\documentstyle[12pt]{article}
\newcommand{\bff}[1]{{\mbox{\boldmath $#1$}}}
\begin{document}
\title{\bf
Monopole giant resonances and nuclear compressibility in                 
relativistic mean field theory}
\author{\\[2.0ex]
D. Vretenar\\
Physics Department, Faculty of Science, University of Zagreb,\\
Zagreb, Croatia\\[2.0ex]
G.\,A. Lalazissis\\
Department of Theoretical Physics, Aristotle University of Thessaloniki\\
GR-54006 Thessaloniki, Greece\\[2.0ex]
R. Behnsch, W. P\" oschl and P. Ring \\       
Physik-Department der Technischen Universit\"at M\"unchen,\\
D-85748 Garching, Germany}
\vspace{10mm}
\date{\today}
\maketitle
\vspace{10mm}
\begin{abstract}
Isoscalar and isovector monopole oscillations that
correspond to giant resonances in spherical nuclei are
described in the framework of time-dependent relativistic
mean-field  (RMF) theory. Excitation energies and the
structure of eigenmodes are determined from a Fourier
analysis of dynamical monopole moments and densities. The
generator coordinate method, with generating functions that
are solutions of constrained RMF calculations, is also used
to calculate excitation energies and transition densities
of giant monopole states.  Calculations are performed with
effective interactions which differ in their prediction of
the nuclear matter compression modulus $K_{\rm nm}$. Both
time-dependent and constrained RMF results indicate that
empirical GMR energies are best reproduced by an effective
force with $K_{\rm nm}\approx 270$ MeV.
\end{abstract}

\newpage
\baselineskip = 24pt

\section{Introduction}
The nuclear matter compression modulus $K_{\rm nm}$ is an
important quantity in the description of properties of
nuclei, supernovae explosions, neutron stars, and heavy ion
collisions. In principle the value of $K_{\rm nm}$ can be
extracted from the experimental energies of isoscalar
monopole vibrations (giant monopole resonances GMR) in
nuclei.  In a semi empirical macroscopic approach, the
analysis is based upon a leptodermous expansion of the
compression modulus of a nucleus, analogous to the liquid
drop mass formula~\cite{TKB.81}.  In principle such an
expansion provides a ``model independent'' determination of
$K_{\rm nm}$. However, the macroscopic approach presents
several ambiguities. The formula itself is based on the
assumption that the breathing mode is a small amplitude
vibration. More important, to correctly interpret the value
of the volume term in the expansion, one has to assume a
certain mode of vibration. It has been argued that a direct
determination of the various contributions to the
compression modulus through a fit of the breathing mode
frequencies cannot provide an accurate value for $K_{\rm
nm}$.  Recently it was shown by Shlomo and
Youngblood~\cite{SY.93} that the complete experimental data
set on isoscalar monopole giant resonances does not limit
the range of $K_{\rm nm}$ to better than $200 - 350$ MeV.
Microscopic calculations of GMR excitation energies might
provide a more reliable approach to the determination of
the nuclear matter compression modulus. One starts by
constructing sets of effective interactions which differ
mostly by their prediction for $K_{\rm nm}$, but otherwise
reproduce reasonably well experimental data on nuclear
properties. Static properties of the ground state are
calculated in the self-consistent mean field approximation,
and RPA calculations are performed for the
excitations~\cite{Bla.80,BBD.95a}. Non relativistic
Hartree-Fock plus RPA calculations, using both Skyrme and
Gogny effective interactions, indicate that the value of
$K_{\rm nm}$ should be in the range 210-220 MeV. The
breathing-mode giant monopole resonances have also been
studied within the framework of the relativistic mean field
(RMF) theory, using the generator coordinate method
(GCM)~\cite{SRS.94}.  It was shown that the GCM succeeds in
describing the GMR energies in nuclei and that the
empirical breathing mode energies of heavy nuclei can be
reproduced by effective forces with $K_{\rm nm}\approx 300$
MeV in the RMF theory.

In the present article we describe isoscalar and isovector
monopole oscillations in spherical nuclei in the framework
of time dependent relativistic mean field theory (RMFT).
The model represents a relativistic generalization of the
time-dependent Hartree-Fock approach.  Nuclear dynamics is
described by the simultaneous evolution of $A$ single
particle wave-functions in the time-dependent mean fields.
Frequencies of eigenmodes are determined from a Fourier
analysis of dynamical quantities. In this microscopic
description, self-consistent mean-field calculations are
performed for static ground-state properties, and
time-dependent calculations for monopole excitations using
the same parameter sets of the Lagrangian.  A basic
advantage of the time dependent model is that no assumption
about the nature of the mode of vibrations has to be made.
Another approach that goes beyond the HF+RPA approximation
is provided by the Generator Coordinate Method. In the
second part of the article we extend the model of Ref.
~\cite{SRS.94}, and calculate the excitation energies of
giant monopole states with relativistic GCM. As generating
functions we use products of Slater determinants , built
from single-particle solutions of constrained RMF
calculations, and coherent states that represent the meson
fields.  

The article is organized as follows. In Sec.~2 we present
the essential features of the time-dependent relativistic
mean-field model, and some details of its application to
spherical nuclei. Results of time-dependent calculations
for a series of doubly closed-shell nuclei and a set of
effective interactions are discussed in Sec.~3. In Sec.~4
details of the relativistic GCM are explained. Sec.~5
contains a discussion of the systematics of energies of
giant monopole states that result from constrained RMF
calculations with GCM. A summary of our results is
presented in Sec.~6.

\section{The time-dependent relativistic mean-field model}

The dynamics of collective motion in nuclei is described in
the framework of relativistic mean-field
theory~\cite{SW.86,Rei.89,Ser.92,GRT.90,Rin.96}. Details of
the time-dependent model are given in
Refs.~\cite{VBR.93,VBR.95}. Here we only outline its
essential features.  In relativistic quantum hadrodynamics
the nucleons, described as Dirac particles, are coupled to
exchange mesons and photons through an effective Lagrangian.
The model is based on the one boson exchange description of
the nucleon-nucleon interaction. The Lagrangian density of 
the model is given as
\begin{eqnarray}
{\cal L}&=&\bar\psi\left(i\gamma\cdot\partial-m\right)\psi
~+~\frac{1}{2}(\partial\sigma)^2-U(\sigma )
\nonumber\\
&&-~\frac{1}{4}\Omega_{\mu\nu}\Omega^{\mu\nu}
+\frac{1}{2}m^2_\omega\omega^2
~-~\frac{1}{4}{\vec{\rm R}}_{\mu\nu}{\vec{\rm R}}^{\mu\nu}
+\frac{1}{2}m^2_\rho\vec\rho^{\,2}
~-~\frac{1}{4}{\rm F}_{\mu\nu}{\rm F}^{\mu\nu}
\nonumber\\
&&-~g_\sigma\bar\psi\sigma\psi~-~
g_\omega\bar\psi\gamma\cdot\omega\psi~-~
g_\rho\bar\psi\gamma\cdot\vec\rho\vec\tau\psi~-~
e\bar\psi\gamma\cdot A \frac{(1-\tau_3)}{2}\psi\;.
\label{lagrangian}
\end{eqnarray}
The Dirac spinor $\psi$ denotes the nucleon with mass $m$.
$m_\sigma$, $m_\omega$, and $m_\rho$ are the masses of the
$\sigma$-meson, the $\omega$-meson, and the $\rho$-meson,
and $g_\sigma$, $g_\omega$, and $g_\rho$ are the
corresponding coupling constants for the mesons to the
nucleon. $U(\sigma)$ denotes the nonlinear $\sigma$
self-interaction
\begin{equation}
U(\sigma)~=~\frac{1}{2}m^2_\sigma\sigma^2+\frac{1}{3}g_2\sigma^3+
\frac{1}{4}g_3\sigma^4,
\label{NL}
\end{equation}
and $\Omega^{\mu\nu}$, $\vec R^{\mu\nu}$, and $F^{\mu\nu}$
are field tensors~\cite{SW.86}.

{}From the Lagrangian density (\ref{lagrangian}) the coupled
equations of motion are derived, the Dirac equation for
the nucleons:
\begin{eqnarray}
i\partial_t\psi_i&=&\left[ \bff\alpha
\left(-i\bff\nabla-g_\omega\bff\omega-
g_\rho\vec\tau\vec{\bff\rho}
-e\frac{(1-\tau_3)}{2}{\bff A}\right)
+\beta(m+g_\sigma \sigma)\right. \nonumber\\
&&\left. +g_\omega \omega_0+g_\rho\vec\tau\vec\rho_0
+e\frac{(1-\tau_3)}{2} A_0
\right]\psi_i
\label{dirac}
\end{eqnarray}
and the Klein-Gordon equations for the mesons:
\begin{eqnarray}
\left(\partial_t^2-\Delta+m^2_\sigma\right)\sigma&=&
-g_\sigma\rho_s-g_2 \sigma^2-g_3 \sigma^3\\
\left(\partial_t^2-\Delta+m^2_\omega\right)\omega_\mu&=&
~g_\omega j_\mu\\
\left(\partial_t^2-\Delta+m^2_\rho\right)\vec\rho_\mu&=&
~g_\rho \vec j_\mu\\
\left(\partial_t^2-\Delta\right)A_\mu&=&
~e j_\mu^{\rm em}.
\label{KGeq4}
\end{eqnarray}
In the relativistic mean field approximation the $A$
nucleons, described by a Slater determinant $|\Phi\rangle$
of single-particle spinors $\psi_i,~(i=1,2,...,A)$, move
independently in the classical meson fields.  The sources
of the fields are calculated in the {\it no-sea}
approximation~\cite{VBR.95}:\
-~the scalar density
\begin{equation}
\rho_{\rm s}~=~\sum_{i=1}^A \bar\psi_i\psi_i,
\label{rho}
\end{equation}
-~the isoscalar baryon current
\begin{equation}
j^\mu~=~\sum_{i=1}^A \bar\psi_i\gamma^\mu\psi_i,
\label{current}
\end{equation}
-~the isovector baryon current
\begin{equation}
\vec j^{\,\mu}~=~\sum_{i=1}^A \bar\psi_i\gamma^\mu \vec \tau\psi_i,
\label{isocurrent}
\end{equation}
-~the electromagnetic current for the photon-field
\begin{equation}
j^\mu_{\rm em}~=~\sum_{i=1}^A
\bar\psi_i\gamma^\mu\frac{1-\tau_3}{2}\psi_i,
\label{emcurrent}
\end{equation}
where the summation is over all occupied states in the
Slater determinant $|\Phi\rangle$. Negative-energy states
do not contribute to the densities in the {\it no-sea}
approximation of the stationary solutions.  However,
negative energy contributions are included implicitly in
the time-dependent calculation, since the Dirac equation is
solved at each step in time for a different basis set.
Negative energy components with respect to the original
ground state basis are taken into account automatically,
even if at each time the {\it no-sea} approximation is
applied.  It is also assumed that nucleon single-particle
states do not mix isospin. Because of charge conservation,
only the 3-component of the isovector $\vec{\rho}$
contributes.

The ground state of a nucleus is described by the
stationary solution of the coupled system of equations
(\ref{dirac})--(\ref{KGeq4}). It specifies part of the
initial conditions for the time-dependent problem. In the
present work we consider doubly closed-shell nuclei, i.e.
nuclei that are spherical in the ground state. The nucleon
spinor is in this case characterized by the angular
momentum $j$, its $z$-projection $m$, the parity $\pi$, and
the isospin $t_3=\pm\frac{1}{2}$ for neutrons and
protons~\cite{GRT.90}
\begin{equation}
\psi({\bf r},t,s,t_3) =
\frac{1}{r}
\left(
\begin{array}{c}
\ \ f(r) \Phi_{ljm} (\theta, \varphi, s) \\
  ig(r) \Phi_{\tilde ljm} (\theta, \varphi, s)
\end{array}
\right)
\, e^{-iEt} \, \chi_\tau(t_3).
\label{wavefunction}
\end{equation}
$\chi_\tau$ is the isospin function, the orbital angular
momenta $l$ and $\tilde l$ are determined by $j$ and the
parity $\pi$, $f(r)$ and $g(r)$ are radial functions, and
$\Phi_{ljm}$ is the tensor product of the orbital and spin
functions
\begin{equation}
\Phi_{ljm}(\theta, \varphi, s)~=~\sum_{m_s m_l} < \frac{1}{2}~m_s
~l~m_l | j~m> Y_{l m_l}(\theta, \varphi)~\chi_{m_s}(s).
\end{equation}

For a given set of initial conditions, i.e. initial values
for the densities and currents in Eqs.
(\ref{rho}--\ref{emcurrent}), the model describes the time
evolution of $A$ single particle wave-functions in the
time-dependent mean fields..  The description of nuclear
dynamics as a time-dependent initial-value problem is
intrinsically semi-classical, since there is no systematic
procedure to derive the initial conditions that
characterize the motion of a specific mode of the nuclear
system.  In principle the theory is quantized by the
requirement that there exist time-periodic solutions of the
equations of motion, which give integer multiples of
Planck's constant for the classical action along one period
~\cite{RVP.96}.  For giant resonances the time-dependence
of collective dynamical quantities is actually not
periodic, since generally giant resonances are not
stationary states of the mean-field Hamiltonian. The
coupling of the mean-field to the particle continuum allows
for the decay of giant resonances by direct escape of
particles.  In the limit of small amplitude oscillations,
however, the energy obtained from the frequency of the
oscillation coincides with the excitation energy of the
collective state. In Refs.~\cite{VBR.95,PVR.96,RVP.96} we
have studied oscillations of isovector dipole, isovector
spin-dipole, isoscalar quadrupole, and isovector quadrupole
character, which correspond to giant resonances in
spherical nuclei.  In Ref.~\cite{RVP.96} double giant
isovector dipole and double giant isoscalar quadrupole
resonances have been described in the framework of the
time-dependent RMF theory and compared with recent
experimental results.  The model reproduces reasonably well
the experimental data on energies and, for light nuclei,
the widths of giant resonances. All the results are
obtained using parameter sets that reproduce ground-state
properties of nuclei, i.e.  no new parameters are
introduced in the model to specifically describe giant
resonances. In the present study we apply the model to
isoscalar and isovector monopole oscillations in spherical
nuclei.  In this microscopic description, self-consistent
mean-field calculations are performed for static
ground-state properties, and time-dependent calculations
for monopole excitations. Using sets of effective
interactions which differ mostly by their prediction of
nuclear matter compressibility, and which otherwise provide
reasonable results for ground state properties, we
calculate the excitation energies of monopole resonances in
a series of spherical nuclei.

In order to excite monopole oscillations in a doubly
closed-shell nucleus, the spherical solution for the
ground-state has to be initially compressed or radially
expanded by scaling the radial coordinate. The amplitudes
$f^{\rm mon}$ and $g^{\rm mon}$ of the Dirac spinor are
defined
\begin{equation}
f^{\rm mon}(r_{\rm mon})  =
(1+a)~f(r),~~~
g^{\rm mon}(r_{\rm mon})  =
(1+a)~g(r)\;.
\label{def}
\end{equation}
The new coordinates are
\begin{equation}
r_{\rm mon}= (1+a)~r.
\end{equation}
A reasonable choice for the parameter is $|a|\approx
0.05-0.1$. The energy transferred in this way to the ground
state of the nucleus is somewhat above the giant resonance
energy, and the resulting oscillations do not show
anharmonicities associated with large amplitude motion.
For the case of isoscalar oscillations the monopole
deformations of the proton and neutron densities have the
same sign. To excite isovector oscillations, the initial
monopole deformation parameters of protons and neutrons
must have opposite signs.  After the initial monopole
deformation (\ref{def}), the proton and neutron densities
have to be normalized.  It should also be emphasized that
apart from the fact that we concentrate on monopole
excitations no assumption about the radial nature of the
mode of vibrations is made in the time-dependent
calculation.  We do not have to assume that the motion is
adiabatic, nor that the mode corresponds to a scaling of the
density.  The frequency dependence of dynamical quantities
and the transition densities are used to determine the
structure of the eigenmodes.  The frequency can be simply
related to nuclear compressibility only if a single
compression mode dominates.

For the case of spherical symmetry, the time-dependent
Dirac equation (\ref{dirac}) reduces to a set of coupled
first-order partial differential equations for the complex
amplitudes $f$ and $g$ of proton and neutron states
\begin{eqnarray}
i\partial_t f&=&(V_0 + g_\sigma \sigma)f+
(\partial_{r}
- {\kappa\over r}- iV_r)g\\
i\partial_t g&=&(V_0-g_\sigma \sigma -2m)g-
(\partial_{r}
+ {\kappa\over r}- iV_r)f,
\end{eqnarray}
where $\kappa = \pm (j+ {1\over 2})$ for $j = l \mp {1\over
2}$, and the indices $0$ and $r$ denote the time and radial
components of the vector field
\begin{equation}
V^\mu= g_\omega \omega^\mu + g_\rho \rho_3^\mu \tau_3 +
e{(1-\tau_3)\over 2}A^\mu.
\end{equation}
For a given set of initial conditions, the equations of
motion propagate the nuclear system in time. The potentials
are solutions of the Klein-Gordon equations
\begin{equation}
\left(-{\partial^2 \over {\partial r^2}} -
{2\over r}{\partial \over {\partial r}}
+m^2_\phi\right) \phi(r)~=~s_\phi (r),
\end{equation}
$m_\phi$ are meson masses for $\phi = \sigma$, $\omega$ and
$\rho$, and zero for the photon.  The source terms are
calculated from~(\ref{rho})--(\ref{emcurrent}) using in
each time step the latest values for the nucleon
amplitudes.  Retardation effects for the meson fields are
not included, i.e. the time derivatives $\partial_t^2$ in
the equations of motions for the meson fields are
neglected. This is justified by the large masses in the
meson propagators causing a short range of the
corresponding meson exchange forces. At each step in time
the meson fields and electromagnetic potentials are
calculated from
\begin{equation}
\phi(r)~=~\int_0^\infty G_\phi (r,r^\prime) s_\phi(r^\prime)
r^{\prime 2} dr^\prime\;,
\end{equation}
where for massive fields
\begin{equation}
G_\phi(r,r^\prime)~=~{1\over {2m_\phi}} {1\over{r r^\prime}}
\left( e^{m_\phi |r-r^\prime|} - e^{-m_\phi|r+r^\prime|}\right),
\end{equation}
and for the Coulomb field
\begin{eqnarray}
G_C(r,r^\prime)&=&
\frac {1}{r}  ~~~~~\mbox{for}~~ r > r^\prime \nonumber\\
G_C(r,r^\prime)&=&
\frac{1}{r^\prime}  ~~~~~\mbox{for}~~ r <  r^\prime.
\end{eqnarray}
The dynamical variables that characterize vibrations of a
nucleus are defined as expectation values of
single-particle operators in the time-dependent Slater
determinant $|\Phi(t)\rangle$ of occupied states.  In the
framework of the TDRMF model the wave-function of the
nuclear system is a Slater determinant at all times.  For
isoscalar monopole vibrations, the time-dependent monopole
moment is simply defined as:
\begin{equation}
\langle r^2(t)\rangle ~=~\frac {1}{A}\langle \Phi(t) |r^2 |\Phi(t)\rangle.
\end{equation}
The corresponding Fourier transform determines the
frequencies of eigenmodes. The transition density for a
specific mode of oscillations is defined as the Fourier
transform of the time-dependent baryon density at the
corresponding frequency. The effective compression
modulus $K_A$ for a nucleus of mass number $A$ is defined
as
\begin{equation}
E_0= \sqrt{{ \hbar^2 A K_A} \over { m < r^2>_0}},
\label{modulus}
\end{equation}
where $E_0$ is the energy of the isoscalar giant monopole
resonance, $m$ is the nucleon mass, and $<r^2>_0$ denotes
the expectation value of $r^2$ in the unperturbed ground
state.  As in Ref.~\cite{BBD.95a}, Eq. (\ref{modulus})
represents just a convenient parameterization of the
dominant $A$-dependence for the energy of the giant
monopole state.

\section{Monopole oscillations in spherical nuclei }

The study of isoscalar monopole resonances in nuclei
provides an important source of information on the nuclear
matter compressibility. We first consider three spherical
nuclei from different regions of the periodic table:
$^{40}$Ca, $^{90}$Zr and $^{208}$Pb.  These nuclei differ
not only in their masses, but also in the ratio of proton
to neutron number. The complete experimental data set on
isoscalar monopole giant resonances has been recently
analyzed by Shlomo and Youngblood~\cite{SY.93}.  Using a
semi empirical macroscopic approach to deduce from the
systematics of GMR the systematics of nuclear
compressibility, and to extrapolate the data to infinite
nuclear matter, they have shown that the complete data set
does not limit the range of $K_{\rm nm}$ to better than
$200 - 350$ MeV. The GMR energy in $^{208}$Pb is rather
well established at $13.7\pm 0.3$ MeV. The average GMR
energy for $^{90}$Zr as deduced from various experiments is
around $16.5$ MeV, and for $^{40}$Ca the value of the
excitation energy adopted in the calculation was $18\pm 1$
MeV~\cite{SY.93}. In light nuclei, of course, not all
monopole strength is seen. 

We have performed time-dependent relativistic mean-field
calculations for six sets of Lagrangian parameters: 
NL1~\cite{RRM.86},
NL3~\cite{LKR.96}, NL-SH~\cite{SNR.93}, NL2~\cite{LFB.86},
HS~\cite{HS.81}, and L1~\cite{TGB.83}  
(in order of increasing value of the nuclear matter compression
modulus $K_{\rm nm}$).
The values of the
parameters are given in Table 1. These effective forces
have been frequently used to calculate properties of
nuclear matter and of finite nuclei.  In this respect our
work parallels the generator coordinate calculations of
giant monopole resonances and the constrained compressibility
within the relativistic mean-field model of
Ref.~\cite{SRS.94}, and the non-relativistic
self-consistent mean-field plus RPA calculation of nuclear
compressibility with Gogny effective interactions of
Ref.~\cite{BBD.95a}.  The idea is to restrict the possible
values of the nuclear matter compression modulus, on the
basis of the excitation energies of giant monopole states
calculated with different effective interactions. In
addition to the four non-linear sets NL1, NL3, NL-SH and
NL2, we also include two older linear parameterizations HS
and L1, which do not include the self-coupling of the sigma
field. The sets NL1, NL-SH and NL2 have been extensively
used in the description of properties of finite nuclei 
\cite{GRT.90,Rin.96}. 
In order to bridge the gap between NL1 ($K_{\rm nm} = 211.7$
MeV), and NL-SH ($K_{\rm nm} = 355.0$ MeV), we have also
included a new effective interaction NL3 ($K_{\rm nm} =
271.8$ MeV).  This new parameter set has been derived
recently~\cite{LKR.96} by fitting ground state properties
of a large number of spherical nuclei, as well as nuclear
matter properties.  It appears that the NL3 effective
interaction reproduces experimental data better than, for
instance, NL1 or NL-SH.

Results of model calculations for $^{40}$Ca are displayed
in Figs.~1a and 1b: time-dependent monopole moments and the
corresponding Fourier power spectra are shown. Time is
measured in units of ${\rm fm/c}$, and the energy $E=\hbar
\omega$ in MeV.  The numerical accuracy is $\Delta E = 2\pi
\hbar c / T_{\rm final} = 2\pi \hbar c / 1000 {\rm fm} \approx
1.2$ MeV.  The frequency of the isoscalar monopole
oscillations generally increases with the nuclear matter
compression modulus, and the vibrations become more
anharmonic. For the first three sets of parameters (NL1,
NL3, and NL-SH) a single mode dominates. For NL2, HS and L1
the Fourier spectra are fragmented.  The NL1 effective
interaction reproduces especially well the ground-state
properties of nuclei close to the stability
line~\cite{GRT.90}, and it appears that the frequency of
monopole oscillations for the NL1 parameter set is very
close to the expected experimental excitation energy of
$18\pm 1$ MeV. 

In Figs.~2a and 2b the results for $^{90}$Zr are shown. The
frequencies increase with $K_{\rm nm}$, the motion is more
anharmonic and more damped. The damping of the monopole
moments comes from the coupling to the continuum, and there
is also a contribution from the damping via the mean-field
(Landau damping). The experimental GMR excitation energy
for $^{90}$Zr of 16.5 MeV is found between the values
calculated for the NL1 and NL3 parameter sets. For NL1
(15.6 MeV), NL3 (19.4 MeV), NL-SH (21.1 MeV) and the lower
frequency of NL2 (21.6 MeV) we display in Fig.~3 the
corresponding transition densities. For NL1, NL3 and NL-SH
the transition densities show a radial dependence
characteristic for the ``breathing'' mode: a change of the
density in the volume at the expense of that on the
surface. The details depend on the underlying shell
structure.  An unusual radial dependence is found for the
NL2 parameterization.  The transition density, both for
protons and neutrons, has a minimum at $r=0$, of the same
sign as on the surface. The monopole oscillations at the
center of the nucleus are in phase with surface
oscillations.  It seems that a similar behavior is found
for HS and L1, although due to fragmentation it is
difficult to decide at which frequency to plot the
transition density.

In Ref.~\cite{BBD.95a} it was stressed that, rather than on
the systematics over the whole periodic table, the
determination of the nuclear compressibility relies more on
a good measurement and microscopic calculations of GMR in a
single heavy nucleus such as $^{208}$Pb. The results of
TDRMF calculations for $^{208}$Pb are presented in Figs.
4-7. In Figs.~4a and 4b we display the monopole moments and
their Fourier power spectra. As one would expect for a
heavy nucleus, there is very little fragmentation and a
single mode dominates for all parameter sets. The
experimental excitation energy $13.7\pm 0.3$ MeV is very
close to the frequency of oscillations obtained with the
NL3 parameter set: 14.1 MeV.  As in the case of $^{90}$Zr,
the calculated excitation energy for the NL1 parameter set
($K_{\rm nm} =211.7$ MeV), is approximately 1 MeV lower
than the average experimental value.

The effective compression modulus $K_A$ (\ref{modulus}) of
$^{208}$Pb is displayed in Fig.~5 as a function of $K_{\rm
nm}$. The observed behavior of $K_A$ is almost identical to
the constrained compressibility calculated in the
relativistic mean-field model using the generator
coordinate method~\cite{SRS.94}. The deviation from the
almost linear behavior, which is observed for the HS linear
parameter set, is slightly more pronounced in the present
time-dependent calculation.

The transition densities that correspond to the main peaks
calculated for the six effective interactions are displayed
in Figs.~6a and 6b. In the left columns we plot the
dynamical transition densities, calculated as Fourier
transforms of the time-dependent baryon densities. On the
right hand sides the scaling densities are
shown~\cite{Bla.80}
\begin{equation}
\rho^{p(n)}_S (r) = 3 \rho^{p(n)}_0 + r {d \over {dr}} \rho^{p(n)}_0 (r),
\end{equation}
where $\rho^{p(n)}_0$ denotes the ground-state vector
density for protons (neutrons). In the scaling model the
transition densities follow from a simple radial scaling of
the ground-state density, both the central density and the
surface thickness vary. We notice that the scaling
transition densities are almost identical, with a possible
exception for L1, for all effective interactions. They do
not provide any information about the dynamics of isoscalar
monopole vibrations. On the other hand, the radial
shape of the transition densities that result from
time-dependent calculations depends very much on the value
of the nuclear matter compression modulus.  As we have
already seen for $^{90}$Zr, starting with NL2 ($K_{\rm nm}
= 399.2$ MeV), a minimum develops in the center of the
nucleus. It also appears that the oscillations of proton
and neutron densities are out of phase at $r=0$. Thus not
only does the frequency of oscillations increase with
$K_{\rm nm}$, but also the radial dependence of the density
oscillations changes dramatically. 

In order to understand better this behavior, we separate 
in Fig.~7 the volume and surface contributions to the
transition densities. Following the procedure of
Ref.~\cite{Bla.80} we define a velocity field associated
with the collective motion
\begin{equation}
{\bf u}({\bf r})~=~- \frac{{\bf r}}{r^3 \rho_0 (r)} \int_0^r r^{\prime 2}
\rho_T(r^\prime, E_0) dr^\prime
\end{equation}
where $\rho_0$ is the ground-state density, and
$\rho_T(r^\prime, E_0)$ denotes the transition densities
shown in Fig.~6. The transition density is separated into
two components
\begin{equation}
\rho_T^{\rm vol}(r, E_0)~=~\rho_0 \bff\nabla{\bf u}~=~
\rho_0 \frac{1}{r^2} \frac {d}{dr} (r^2 u)
\end{equation}
\begin{equation}
\rho_T^{\rm surf}(r, E_0) = {\bf u}\bff\nabla \rho_0  =
u \frac {d\rho_0}{dr}.
\end{equation}
The resulting volume and surface transition densities are
shown in Fig.~7 for all six effective interactions. We
notice that the surface contribution does not depend much
on the parameter set used, that is, on the nuclear matter
incompressibility. The volume transition density, as one
would expect, is very sensitive to the value of $K_{\rm
nm}$.  A very interesting phenomenon is the formation of
standing waves in the bulk. It starts already for NL2, but
is clearly observed for HS and L1.

The effective interactions NL1 and NL3 seem to produce GMR
excitation energies which are quite close to the
experimental values. We have therefore calculated, for
these two parameter sets, the isoscalar giant monopole
resonances in a number of doubly closed-shell nuclei:
$^{56}$Ni, $^{100,132}$Sn, $^{122}$Zr, $^{146}$Gd. The
results, together with those already discussed, are shown
in Fig.~8. The energies of giant monopole states are
determined from the Fourier spectra of the time-dependent
monopole moments, and are displayed as function of the mass
number. We notice that the NL1 excitation energies are
systematically lower, but that otherwise the two effective
interactions produce very similar dependence on the mass
number. The empirical curve $E_x \approx 80~A^{-1/3}$ MeV
is also included in the figure, and it follows very closely
the excitation energies calculated with the NL3 parameter
set.

Experimental data on isovector giant monopole resonances
are much less known. The systematics of excitation energies
will not, in general, depend on the nuclear matter
compression modulus. More likely, energies will depend on
the coefficient of asymmetry energy.  Isovector excitations
are therefore outside the main topic of the present study.
Nevertheless we have calculated the eigenfrequencies of
isovector monopole modes, and compared them with available
data on energies of giant monopole resonances.  The
time-dependent isovector monopole moments
\begin{displaymath}
< r^2_{\rm p} > - < r^2_{\rm n} >
\end{displaymath}
and the corresponding Fourier spectra for $^{208}$Pb are
displayed in Fig.~9. Compared to the isoscalar vibrations
(Fig.~4), we notice that the isovector oscillations are
more damped and the anharmonicities are more pronounced.
Consequently, the Fourier spectra are more fragmented.  In
each figure only the energy of the main peak is displayed.
The four non-linear parameter sets produce similar Fourier
spectra, with most of the strength between 25 MeV and 30
MeV. The fragmentation of the Fourier spectra is more
pronounced for the linear effective interactions HS and L1,
with a considerable amount of strength shifted in the
energy region $30 - 40$ MeV. The results of time-dependent
calculations should be compared with the experimental value
for the IV GMR in $^{208}$Pb: $26\pm 3$ MeV~\cite{GBM.90}.
We have also calculated the time-dependent isovector
monopole moments for $^{40}$Ca and $^{90}$Zr. Although the
oscillations are more anharmonic as compared to $^{208}$Pb,
the corresponding Fourier spectra are generally in
agreement with the experimental data on the IV GMR:
31.1$\pm$2.2 MeV for $^{40}$Ca and 28.5$\pm$2.6 MeV for
$^{90}$Zr~\cite{GBM.90}.

\section{Constrained relativistic mean-field calculations}

A very useful method for description of excited states in
nuclei is provided by constrained mean-field calculations.
In the framework of the relativistic mean-field theory,
constrained calculations were performed for the ground
state of $^{24}$Mg, using two quadratic
constraints~\cite{KR.88}.  By analyzing the curvature of
the energy surface near the ground state, in
Ref.~\cite{SCR.94} constrained RMF calculations were used
to determine the dependence of the excitation energy of
giant monopole states on the nuclear compressibility.  In
Ref.~\cite{SRS.94} this approach has been generalized by
applying the generator coordinate method (GCM) to the RMF.
The GCM takes into account correlations produced by
collective motion of the nucleons.

In the present study we further extend the method of
Ref.~\cite{SRS.94}, by using a more general ansatz for the
generating functions of the GCM. Specifically, in
Ref.~\cite{SRS.94} Slater determinants of nucleon
single-particle spinors, resulting from constrained RMF
calculations, were used as generating functions. Here the
bosonic sector is explicitly taken into account. The
generating functions will be defined as direct products of
Slater determinants built from single-particle spinors and
the coherent states that represent the meson fields.
Furthermore, the investigation is extended to include the
isovector ($T=1$) giant monopole states.

The GCM $A$-particle trial wave-function $\Psi_{\rm
GCM}({\bf r}_1,\dots,{\bf r}_A)$ is written in the form
\begin{equation}
\Psi_{\rm GCM}({\bf r}_1,\dots,{\bf r}_A) = 
\int f(q)\,\Psi ({\bf r}_1,\dots,{\bf r}_A;q)\,dq\;,
\label{hw-wf}
\end{equation}
where $\Psi({\bf r}_1,\dots,{\bf r}_A;q)$ are the
``generating functions'', and $f(q)$ is the ``generator''
or ``weight function'' that depends on the ``generator
coordinate'' $q$. From the variation of the energy of the
system with respect to $f(q)$, the {\em Hill-Wheeler}
integral equation~\cite{HW.53} for the weight function
$f(q)$ is derived
\begin{equation}
\int \left [ {\cal H}(q,q^{\prime }) - E{\cal N}(q,q^{\prime })\right] 
 f(q^{\prime })\,dq^{\prime } = 0 \; , 
\label{hw-eq}
\end{equation}
where
\begin{equation} 
{\cal H}(q,q^{\prime }) = \langle\Psi(q)|\hat{H}|\Psi(q^\prime)\rangle, 
\label{hw-h}
\end{equation} 
and
\begin{equation} 
{\cal N}(q,q^{\prime }) = \langle\Psi(q)|\Psi(q^\prime)\rangle, 
\label{hw-n}
\end{equation}
are the Hamiltonian and normalization kernels,
respectively. The solutions of the Hill-Wheeler equation
are the discrete eigenvalues $E_n$ and eigenfunctions
$f_n(q)$, which determine the nuclear wave-functions for
the ground and excited states (for details see Ref.
\cite{RS.80}).

The relativistic extension of the GCM uses the
self-consistent solutions of the equations of motion of the
Lagrangian density (\ref{lagrangian}).  The generating
functions are defined as products of Slater determinants
$\Phi(q)$, built from constrained RMF solutions for the
Dirac single-nucleon spinors $\psi_i({\bf r};q)$, and the
coherent states of the $\sigma$, $\omega$, and $\rho$ meson
field and the electromagnetic field
\begin{equation}
|\Psi(q)\rangle=|\Phi(q)\rangle\otimes|\sigma(q)\rangle\otimes
        |\omega_\mu(q)\rangle
        \otimes|\vec \rho_\mu(q)\rangle\otimes|A_\mu(q)\rangle\;.            
\label{state}
\end{equation}
The general expression for the boson coherent state
reads~\cite{IZ.80}:
\begin{equation}
|\phi_\mu\rangle~=~C^{-\frac{1}{2}}
 \exp \left[\int d^3 \tilde{k}\,\omega(k) 
 \phi_\mu({\bf k}) a^+({\bf k}) \right] |0\rangle\;,
\label{coh}
\end{equation}
where $\phi_\mu$ denotes the $\sigma$, $\omega_\mu$,
$\rho_\mu$ and $A_\mu$ fields, $\phi_\mu({\bf k})$ is the
Fourier-transform of the field $\phi_\mu({\bf r})$,
\begin{equation}
\phi_\mu({\bf k})~=~\omega(k) \int d^3r \,\phi_\mu({\bf r})
  e^{i{\bf kr}}\;,
\label{phi-int}
\end{equation}
\begin{equation}
\omega(k)~=~k_0~=~\sqrt{{\bf k}^2 + m_{\phi_\mu}^2}\;,
\label{omega}
\end{equation}
and 
\begin{equation}
d^3 \tilde{k}~=~{{d^3k} \over {(2\pi)^3 2\omega(k)}}~=~ 
\frac{d^4 k}{(2\pi)^4} (2\pi) 
\delta (k \cdot k - m_{\phi_\mu}^2)\,\theta(k^0).
\label{k-tild}
\end{equation}
$a^+({\bf k})$ is the creation operator of a boson
with momentum $\bf k$ and $C$ is the normalization
constant.  Onishi's formulas~\cite{RS.80} are used in the
calculation of the integral kernels (\ref{hw-h}) and
(\ref{hw-n}) in the solution of the Hill-Wheeler equation.
The normalization kernel can be written as product of
fermionic and bosonic factors
\begin{equation}
{\cal N}(q,q^\prime)~\equiv~
{\cal N}_{\rm D}(q,q^\prime)
{\cal N}_{\rm M}(q,q^\prime),
\end{equation}
where
\begin{equation}
{\cal N}_{\rm D}(q,q^{\prime })~\equiv~
\langle\Phi(q)|\Phi(q^\prime)\rangle 
~=~\det\{N_{ij}(q,q^\prime)\}, 
\label{hw-nnd}
\end{equation}
with the matrix
\begin{equation}
N_{ij}(q,q^\prime)~=~\langle \psi_i(q)|\psi_j(q^\prime)\rangle,  
\label{sp-overlaps}
\end{equation}
where $|\psi_i(q)\rangle$ denotes the single-particle
spinors. The boson factor
\begin{equation}
{\cal N}_{\rm M}(q,q^\prime)~=~
\langle \sigma(q)|\sigma(q^\prime)\rangle \;
\langle \omega_\mu(q)|\omega_\mu(q^\prime)\rangle \;
\langle \vec \rho_\mu(q)|\vec \rho_\mu(q^\prime)\rangle \;
\langle A_\mu(q)|A_\mu(q^\prime)\rangle
\label{hw-nnm}
\end{equation}
can be calculated from the overlap kernel
\begin{equation}
\langle\phi_\mu(q)|\phi_\mu(q^\prime)\rangle~=~
\exp \left( - \frac{1}{2} \int d^3 \tilde k\, 
|\phi_\mu({\bf k};q)-{\phi}_\mu({\bf k};q^\prime)|^2 \right).
\label{phi-ov}
\end{equation}
The Hamiltonian kernel ${\cal H}(q,q^{\prime })$ reads 
\begin{equation} 
{\cal H}(q,q^{\prime })~=~
{\cal N}(q,q^{\prime }) \int {\cal H }(r;q,q^\prime)\,dr,
\label{hw-hh}
\end{equation} 
where, for the case of spherical symmetry, the energy
density kernel ${\cal H} (r;q,q^\prime)$ is given as
\begin{eqnarray}
{\cal H }_{\rm RMF}(r;q,q^\prime)&=&
\sum_{i,j=1}^A N^{-1}_{ji}\,
\psi_i^+(r;q)\{-i\bff{\alpha\nabla}\}\psi_j(r;q^\prime)
 + M \rho_s(r;q,q^\prime) 
\label{hw-hq}
\nonumber\\
&&{}+\frac{1}{2} g_\sigma\rho_s(r;q,q^\prime)\sigma(r;q,q^\prime) 
 + \frac{1}{2} g_\omega \rho_v(r;q,q^\prime) \omega^0(r;q,q^\prime) 
\nonumber\\
&&{}+\frac{1}{2} g_\rho \rho_3(r;q,q^\prime)\rho^0(r;q,q^\prime) 
 + \frac{1}{2} e \rho_{\rm p}(r;q,q^\prime) A^0(r;q,q^\prime) 
\nonumber\\
&&{}+\frac{1}{2}((\bff\nabla\sigma(r;q,q^\prime))^2+
    m_\sigma^2\sigma^2(r;q,q^\prime) )
+\frac{1}{3}g_2 \sigma^3+\frac{1}{4}g_3 \sigma^4 
\nonumber\\
&&{} - \frac{1}{2}((\bff\nabla\omega^0(r;q,q^\prime))^2+
    m_\omega^2(\omega^0(r;q,q^\prime))^2) 
\nonumber\\ 
&&{}-\frac{1}{2}((\bff\nabla \rho^0(r;q,q^\prime))^2+
    m_\rho^2 (\rho^0(r;q,q^\prime))^2) \nonumber\\
&&{}-\frac{1}{2}(\bff\nabla A^0(r;q,q^\prime))^2,
\end{eqnarray}
with the density matrices 
\begin{eqnarray}
\rho_s(r;q,q^\prime)&=&\sum_{i,j=1}^A N^{-1}_{ji}\,
\bar\psi_i({\bf r};q)\psi_j({\bf r};q^\prime)\\ 
\rho_v(r;q,q^\prime)&=&\sum_{i,j=1}^A N^{-1}_{ji}\,
\psi_i^+({\bf r};q)\psi_j({\bf r};q^\prime)\\
\rho_3(r;q,q^\prime)&=&\sum_{i,j=1}^A N^{-1}_{ji}\,
\psi_i^+({\bf r};q)\tau_3\psi_j({\bf r};q^\prime)\\ 
\rho_{\rm p}(r;q,q^\prime)&=&\sum_{i,j=1}^A N^{-1}_{ji}\,
\psi_i^+({\bf r};q) {{(1-\tau_3)}\over 2} 
\psi_j({\bf r};q^\prime)\;.
\label{densities}
\end{eqnarray}
The densities are sources for the Klein-Gordon equations.
Their solutions are the meson fields that appear in Eq.\
(\ref{hw-hq})
\begin{equation}
 \phi_\mu(r;q,q^\prime)~=~\frac{1}{2} 
(\phi_\mu(r;q) + \phi_\mu(r;q^\prime).
\label{phi}
\end{equation}

\section{Isoscalar and isovector Giant Monopole Resonances}

Solutions of constrained RMF calculations have fixed
expectation value $Q$ of some observable $\hat Q$.  The
constraining operator $\hat Q$ is coupled to the Dirac
Hamiltonian through a Lagrange multiplier $q$
\begin{equation}
\left(\hat h_{\rm D}-q\,\hat Q\right)\psi_i({\bf r};q)~=~
\varepsilon_i \psi_i({\bf r};q),
\label{d-eq-constrained}
\end{equation}
with the Dirac Hamiltonian 
\begin{equation}
\hat h_{\rm D}~=~-i\bff{\alpha\nabla} + 
\beta ( (M+g_\sigma \sigma) +
g_\omega \gamma^\mu \omega_\mu + 
g_\rho \gamma^\mu \vec\tau \vec \rho_\mu + 
e\,\gamma^\mu {{(1-\tau_3)} \over 2} A_\mu ).
\label{d-operator}
\end{equation}
The constrained Dirac equation and the corresponding
Klein-Gordon equations for the fields are solved
self-consistently.  In relativistic GCM calculations the
Lagrange multiplier $q$ is chosen as the generator
coordinate.  The constraint operator reads
\begin{equation}
\hat Q~\equiv~\hat Q^T~=~\sum_{i=1}^A\;(2t_i)^Tr_i^2,
\quad\quad T=0,\,1
\label{c-operator}
\end{equation}
and it relates the generator coordinate to nuclear {\em
rms}\/ radii
\begin{eqnarray}
\langle\,\hat Q\,\rangle_{T=0}&=&\langle\,r^2\,\rangle~=~ 
\frac{1}{A} \int r^2 \rho_v({\bf r};q) d^3 r
\nonumber\\
\langle\,\hat Q\,\rangle_{T=1}&=& 
\langle\,r^2_{\rm n}\,\rangle - \langle\, r^2_{\rm p}\,\rangle~=~ 
\frac{1}{N} \int r^2 \rho^{\rm n}_v({\bf r};q) d^3 r - 
\frac{1}{Z} \int r^2 \rho^{\rm p}_v({\bf r};q) d^3 r.
\end{eqnarray}
Using a Hartree code in coordinate space, constrained RMF
calculations for the spherical symmetric case are performed
for a number of values of the Lagrange multipliers $q$. The
allowed values of $q$ are limited by the requirement that
the coupled system of Dirac and Klein-Gordon equations
converges to a self-consistent solution.  

{}From a set of constrained RMF solutions (single-particle
spinors of occupied states, meson and electromagnetic
fields), the normalization (\ref{hw-n}) and energy
(\ref{hw-h}) kernels are calculated. Solutions of the
associated integral Hill-Wheeler equation determine the
nuclear ground and excited states.  The method of
``symmetrical orthogonalization''~\cite{RS.80,FV.76} is
applied to the generalized eigenvalue problem.  The
Hamiltonian kernel is projected onto the regular
(collective) subspace of the norm kernel.  A series of
collective eigenenergies of the constrained system and the
corresponding nuclear wave-functions are obtained, which
can be interpreted as the correlated ground-state solution
and excited monopole states.  Furthermore,  nuclear
densities in the ground-state and first excited state are
calculated, as well as the transition densities
\begin{equation}
\rho_{fi}(r)~=~
\int dq \, dq' \, f_f(q) f_i(q') {\cal N}(q,q') \rho_v(r;q,q')\;,
\label{trans-dens}
\end{equation}
where $f$ and $i$ denote the resonance and the
ground-state, respectively.  Using the four non-linear
effective interactions NL1, NL3, NL-SH and NL2,
calculations have been performed for a number of spherical
nuclei.
 
In Table 2 the isoscalar monopole excitation energies are
displayed. For the NL1 parameter set we also include in the
second row the excitation energies obtained with
constrained GCM calculations from Ref.~\cite{SRS.94}.  It
appears that the inclusion of meson coherent states in the
generating functions lowers the calculated excitation
energies by approximately 0.5 MeV. Compared to the results
of time-dependent calculations of Sec.~3, the energies in
Table 2 are lower by more than 1 MeV, and is some cases the
difference is larger than 2 MeV. The reason is that in
constrained GCM calculations one implicitly assumes that
the motion is adiabatic. In the time-dependent calculations
on the other hand, no assumption about the nature of the
mode of vibration is made.  The comparison between
constrained and time-dependent mean-field calculations
strongly emphasizes the fact that we have to understand how
the nucleus vibrates in order to relate the excitation
energies of monopole states to the nuclear matter
compression modulus.

In Fig.~10 the calculated energies of giant monopole states
are plotted as functions of $A^{-1/3}$. An average linear
dependence is observed for all four parameter sets. The
deviations from a pure linear dependence are relatively
small for NL1 and NL3.  They are more pronounced for the
effective interactions NL-SH and NL2, for which the nuclear
matter compression modulus $K_{\rm nm} > 300$ MeV. The
effective nuclear compression moduli $K_A$ (\ref{modulus})
of $^{16}$O, $^{40}$Ca, $^{90}$Zr and $^{208}$Pb are shown
in Fig.~11 as functions of $K_{\rm nm}$. The GCM results
for $E_0$ and $<r^2>_0$ are used in (\ref{modulus}).  We
did not calculate $K_A$ from the curvature of the energy
surface, and then $E_0$ from (\ref{modulus}), as was done
in Refs.~\cite{SCR.94,SRS.94}.  The relation
(\ref{modulus}) is used to obtain an estimate for the
values $K_A$ that result from GCM calculation, in which the
dynamics of the system is described by the non diagonal
matrix elements of the energy kernel.  Due to the
correlations that these matrix elements produce, the
calculated values of $K_A$ are somewhat lower than the
corresponding results of Ref.~\cite{SRS.94}. For $^{208}$Pb
the values of $K_A$ can be compared with results of
time-dependent calculations (see Fig.~5).  The constrained
GCM transition densities (\ref{trans-dens}) for $^{208}$Pb
are shown in Fig.~12.  Similar in shape to those calculated
in the scaling model (Fig.~6), the constrained transition
densities display a radial behavior which does not
significantly depend on the effective interaction. In
contrast to the results of time-dependent calculation, the
dynamics of isoscalar monopole vibrations which is
described by the constrained transition densities does not
depend on the effective compression modulus. Again, this is
due to the assumption of adiabatic motion, which is
inherent in the constrained GCM approach.
 
In Table 3 we display the excitation energies of isovector
monopole states in $^{40}$Ca, $^{90}$Zr and $^{208}$Pb, and
compare the theoretical values with experimental data on
isovector giant monopole resonances~\cite{GBM.90}. The
calculations have been performed for the four non-linear
parameter sets, and in the second column the corresponding
asymmetry energy coefficients $a_{\rm sym}$ are included.
We notice that the calculated excitation energies, like the
results of time-dependent calculation, do not depend very
much on the effective interaction.  In particular, no
simple relation can be established between the coefficient
of asymmetry energy and the excitation energies of
isovector states. The calculated values for $^{40}$Ca and
$^{90}$Zr are close to the experimental excitation
energies.  For $^{208}$Pb, however, constrained GCM
calculations produce results which are almost 10 MeV lower
that the experimental IV GMR. This is in contrast with the
results of time-dependent calculation which center around
28 MeV, in agreement with experimental data. This result
might indicate that in a heavy nucleus such as $^{208}$Pb
the assumption of adiabatic motion is not justified for the
isovector mode. On the other hand, the constrained GCM
results might also indicate that some strength should be
expected around 17 MeV for the isovector monopole mode in
$^{208}$Pb. The corresponding time-dependent results (Fig.
9) display some strength below the main peak in the Fourier
power spectra, but only very little below 20 MeV.  Finally,
the isovector transition densities for $^{208}$Pb are shown
Fig.~13. As we have already seen for the isoscalar mode,
the transition densities depend very little on the
effective interactions.

\section{Summary }

Relativistic mean field theory has been used to investigate
the monopole eigenmodes of a number of spherical closed
shell nuclei, from $^{16}$O to
the heavy nucleus $^{208}$Pb. Two microscopic methods have
been applied:

\begin{enumerate}
\item[a)]Time-dependent relativistic mean-field
calculations.
Starting from the self-consistent mean-field 
solution for the ground state and a set of 
appropriate initial conditions, the
full set of time-dependent RMF equations is solved and 
dynamical variables are analyzed.
Collective isoscalar and isovector monopole oscillations are  
studied, which correspond to giant resonances in spherical nuclei.
Excitation energies and the
structure of eigenmodes are determined from a Fourier
analysis of dynamical monopole moments and densities. 

\item[b)] The
generator coordinate method, with generating functions that
are solutions of constrained RMF calculations, is used
to calculate excitation energies and transition densities
of giant monopole states. Constrained RMF calculations,  
with $r^2$ (T=0 and T=1) as the constraint operator,
determine the stationary
wave-functions with different values of nuclear $rms$
radii.
Using the generator
coordinate method, the RMF-Hamiltonian is diagonalized
in the basis of these solutions. The lowest eigenmode
corresponds to the correlated ground state. The solution 
for the first
excited state describes the giant monopole resonance.
\end{enumerate}

Calculations have been performed for six different
parameter sets of the Lagrangian. Two of them correspond
to older effective forces,
which do not include nonlinear self-interaction of the
$\sigma$-field. The others are more modern
parametrizations, which have been used with great success
in recent years for a unified description of 
properties of nuclei over the periodic table
(binding energies, radii, isotopic shifts, deformation
parameters, moments of inertia in superdeformed nuclei, 
nuclear halos at the neutron drip line). The six
parameter sets differ essentially in the deduced value for
the incompressibility of nuclear matter. Therefore, from 
the energy spectra and transition densities
calculated with these effective forces, it has been 
possible to study
the connection between the incompressibility of nuclear
matter and the breathing mode energy of spherical nuclei.

For the isoscalar mode we have found an almost linear
relation between the excitation energy of the monopole
resonance and the nuclear matter compression modulus.
GCM calculations based on
constrained solutions of the RMF equations yield, in general,
slightly lower excitations energies as compared with 
results of time-dependent calculations. This can be
understood from the assumption of adiabatic motion which is
implicitly contained in constrained RMF calculations. No 
assumption about the nature of the mode is made in the 
time-dependent model. With both methods, the effective 
interactions NL1 and NL3 produce GMR excitation energies 
which are close to the experimental data.
Based on the very precise experimental
value for the isoscalar GMR energy in the heavy nucleus
$^{208}$Pb, which is rather well reproduced by the set NL3,
we have derived a value of approximately 270 MeV for the
incompressibility of nuclear matter. For a series of 
doubly closed-shell nuclei, the effective interaction 
NL3 reproduces the empirical result 
$E_x \approx 80~A^{-1/3}$ MeV.

Transition densities provide valuable
information on the structure of the different modes. It
turns out that there are essential differences
between the two methods. The constrained
GCM transition densities are similar to those of
the simple scaling model, and display a radial 
behavior which does not significantly depend on the 
effective force. This is not the case for
the time-dependent RMF calculations. Here the radial shape 
of the transition densities depends very much on the value
of the nuclear matter compression modulus.
Only for the parameter sets NL1, NL3, and NLSH, the transition
densities display a radial shape characteristic for 
the breathing mode.
For the other sets we find essential
deviations in the nuclear interior, sometimes even standing
waves. Only the surface component of the transition density
is rather independent on the effective force.

As expected, for the isovector mode we have found no 
relation between the
excitation energy of the giant resonance
and the calculated incompressibility of
nuclear matter. 
More damping and more fragmentation in the
spectra is observed, compared to the isoscalar case.  
Excitation energies that result from 
time-dependent calculations
are in good agreement with available experimental data. 
The constrained RMF results agree with experiment only for
light and medium heavy nuclei. For $^{208}$Pb we find
results which are almost 10 MeV lower
that the experimental T=1 GMR. 
This might indicate that in heavy nuclei 
the assumption of adiabatic motion is not justified for the
isovector mode. 

Summarizing our investigations in the framework of
relativistic mean field theory, we conclude that 
the nuclear matter compression modulus
$K_{\rm nm} \approx 270$ MeV 
is in reasonable
agreement with the available data on spherical nuclei.

\newpage
\vspace{20mm}
{\bf Acknowledgments}
\vspace{10mm}

This work has been supported in part by the
Bundesministerium f\"ur Bildung und Forschung under contract
06~TM~743 (6). One of the authors (GAL) acknowledges
support from the European Union under contract TMR-EU/ERB
FMBCICT-950216.


\newpage

\leftline{\Large {\bf Figure Captions}}
\parindent = 2 true cm
\begin{description}
\item[Fig.~1a] Time-dependent isoscalar monopole moments $<r^2>(t)$
and the corresponding Fourier power spectra for $^{40}$Ca.
The parameter sets are NL1, NL3 and NL-SH.

\item[Fig.~1b] Same as in Fig.~1a, but for the effective
interactions NL2, HS and L1.

\item[Fig.~2a] Time-dependent isoscalar monopole moments
$<r^2>(t)$ and the corresponding Fourier power spectra for
$^{90}$Zr.  The parameter sets are NL1, NL3 and NL-SH.

\item[Fig.~2b] Same as in Fig.~2a, but for the effective
interactions NL2, HS and L1.

\item[Fig.~3] Transition densities for the isoscalar
monopole states in $^{90}$Zr.

\item[Fig.~4a] Time-dependent isoscalar monopole moments
$<r^2>(t)$ and the corresponding Fourier power spectra for
$^{208}$Pb.  The parameter sets are NL1, NL3 and NL-SH.

\item[Fig.~4b] Same as in Fig.~4a, but for the effective
interactions NL2, HS and L1.

\item[Fig.~5] The compression modulus $K_A$ (\ref{modulus})
of $^{208}$Pb as function of the nuclear matter compression
modulus $K_{\rm nm}$.  The excitation energies result from
time-dependent relativistic mean-field calculations.

\item[Fig.~6a]  Transition densities calculated as Fourier
transforms of time-dependent baryon densities (left
column), and scaling transition densities (right column),
for the isoscalar monopole states in $^{208}$Pb.  The
parameter sets are NL1, NL3 and NL-SH. Solid lines denote
proton densities, and dashed lines correspond to neutron
transition densities.

\item[Fig.~6b] Same as in Fig.~6a, but for the effective
interactions NL2, HS and L1.

\item[Fig.~7] Volume and surface transition densities for
the isoscalar monopole states in $^{208}$Pb.  Solid lines
correspond to proton densities, and dashed lines denote
neutron transition densities.

\item[Fig.~8] Excitation energies of isoscalar giant
monopole resonances in doubly closed-shell nuclei as
function of the mass number. The effective interactions
are: NL1 (squares) and NL3 (circles). The solid curve
corresponds to the empirical relation $\approx 80~A^{-1/3}$
MeV.

\item[Fig.~9a] Time-dependent isovector monopole moments
$<r_{\rm p}^2> - <r_{\rm n}^2>$ and the corresponding
Fourier power spectra for $^{208}$Pb.  The parameter sets
are NL1, NL3 and NL-SH.

\item[Fig.~9b] Same as in Fig.~9a, but for the effective
interactions NL2, HS and L1.

\item[Fig.~10] Constrained relativistic GCM excitation
energies of isoscalar monopole states in doubly
closed-shell nuclei as function of mass number. The
effective interactions are NL1, NL3, NL-SH and NL2.

\item[Fig.~11] The effective nuclear compression modulus
$K_A$~(\ref{modulus}) of $^{16}$O, $^{40}$Ca, $^{90}$Zr and
$^{208}$Pb as function of $K_{\rm nm}$. The excitation
energies of monopole states are calculated with the
constrained relativistic GCM.

\item[Fig.~12] Constrained GCM transition densities for the
isoscalar monopole states in $^{208}$Pb.

\item[Fig.~13] Constrained GCM transition densities for the
isovector monopole states in $^{208}$Pb.
\end{description}

\newpage
\begin{table}
\caption{Parameter sets for the effective Lagrangian.}
\begin{center}
\begin{tabular}{cr@{.}lr@{.}lr@{.}lr@{.}lr@{.}lr@{.}l}
\hline
&  \multicolumn{2}{c}{NL1}&  \multicolumn{2}{c}{NL3}&
  \multicolumn{2}{c}{NL-SH}&  \multicolumn{2}{c}{NL2}& 
  \multicolumn{2}{c}{HS}& \multicolumn{2}{c}{L1}\\ 
\hline
$m$ [MeV]&    938&0& 939&0& 939&0& 938&0& 939&0& 938&0\\
$m_\sigma$ [MeV]&    492&25& 508&194& 526&0592& 504&89& 520&0& 550&0\\
$m_\omega$ [MeV]&    795&335& 782&501& 783&0& 780&0& 783&0& 783&0\\
$m_\rho$ [MeV]&  763&0&  763&0&  763&0& 763&0& 770&0\\
$g_\sigma$&    10&138& 10&217& 10&44355& 9&111& 10&47& 10&30\\
$g_\omega$&    13&285& 12&868& 12&9451& 11&493& 13&80& 12&60\\
$g_\rho$&    4&975& 4&474& 4&3828& 5&507& 4&04\\
$g_2$ [fm$^{-1}$]&    $-$12&172& $-$10&431& $-$6&9099& $-$2&304\\
$g_3$&    $-$36&265& $-$28&885& $-$15&8337& 13&783\\ 
\hline
$K_{\rm nm}$ [MeV]&  211&7& 271&8& 355&0& 399&2& 545&0& 626&3\\
\hline
\end{tabular}
\end{center}
\end{table}

\begin{table}
\caption{Constrained GCM energies of isoscalar monopole
states. The values of $K_{\rm nm}$ and the excitation
energies are in MeV.}
\begin{center}
\begin{tabular}{rlcccccc}
  & & $K_{\rm nm}$ & $^{16}$O & $^{40}$Ca & $^{48}$Ca & $^{90}$Zr
 & $^{208}$Pb \\ \hline \hline
1&NL-1                &211.7& 20.2 & 16.6 & 15.9 & 14.1 & 11.0 \\
2&NL-1 \cite{SRS.94}&211.7& 20.6 & 17.1 &      & 14.7 & 11.7 \\
3&NL-3                &271.8& 22.6 & 19.6 & 18.9 & 16.9 & 13.0 \\
4&NL-SH               &355.0& 25.0 & 22.0 & 21.5 & 19.5 & 15.0 \\
5&NL-2                &399.2& 27.1 & 24.4 & 23.0 & 21.9 & 16.6 \\ \hline
\end{tabular}
\end{center}
\end{table}

\begin{table}
\caption{Constrained GCM energies of isovector monopole states. The values
of $a_{\rm sym}$ and the excitation energies are in MeV.} 
\begin{center}
\begin{tabular}{lcccc}
  & $a_{\rm sym}$ & $^{40}$Ca  & $^{90}$Zr
 & $^{208}$Pb \\ \hline \hline
NL-1                &43.5& 29.0 & 26..3 & 16.5 \\
NL-3                &37.4& 28.6 & 27.4 & 18.0 \\
NL-SH               &36.1& 28.5 & 27.9 & 18.4 \\
NL-2                &43.9& 30.3 & 28.8 & 16.9 \\ \hline
exp.~\cite{GBM.90}&& 31.1$\pm$2.2  & 28.5$\pm$2.6 & 26.0$\pm$3.0\\ \hline 
\end{tabular}
\end{center}
\end{table}

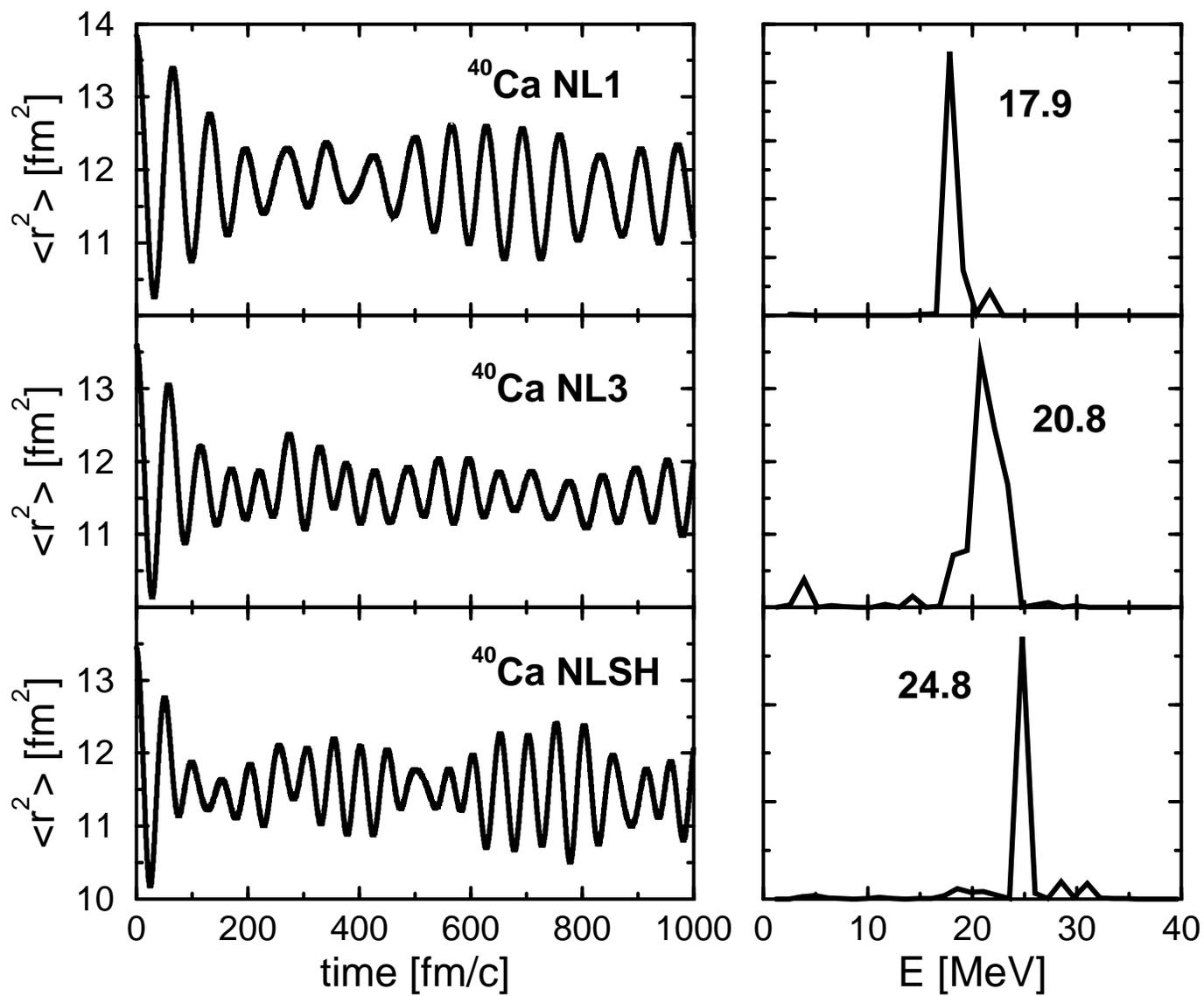
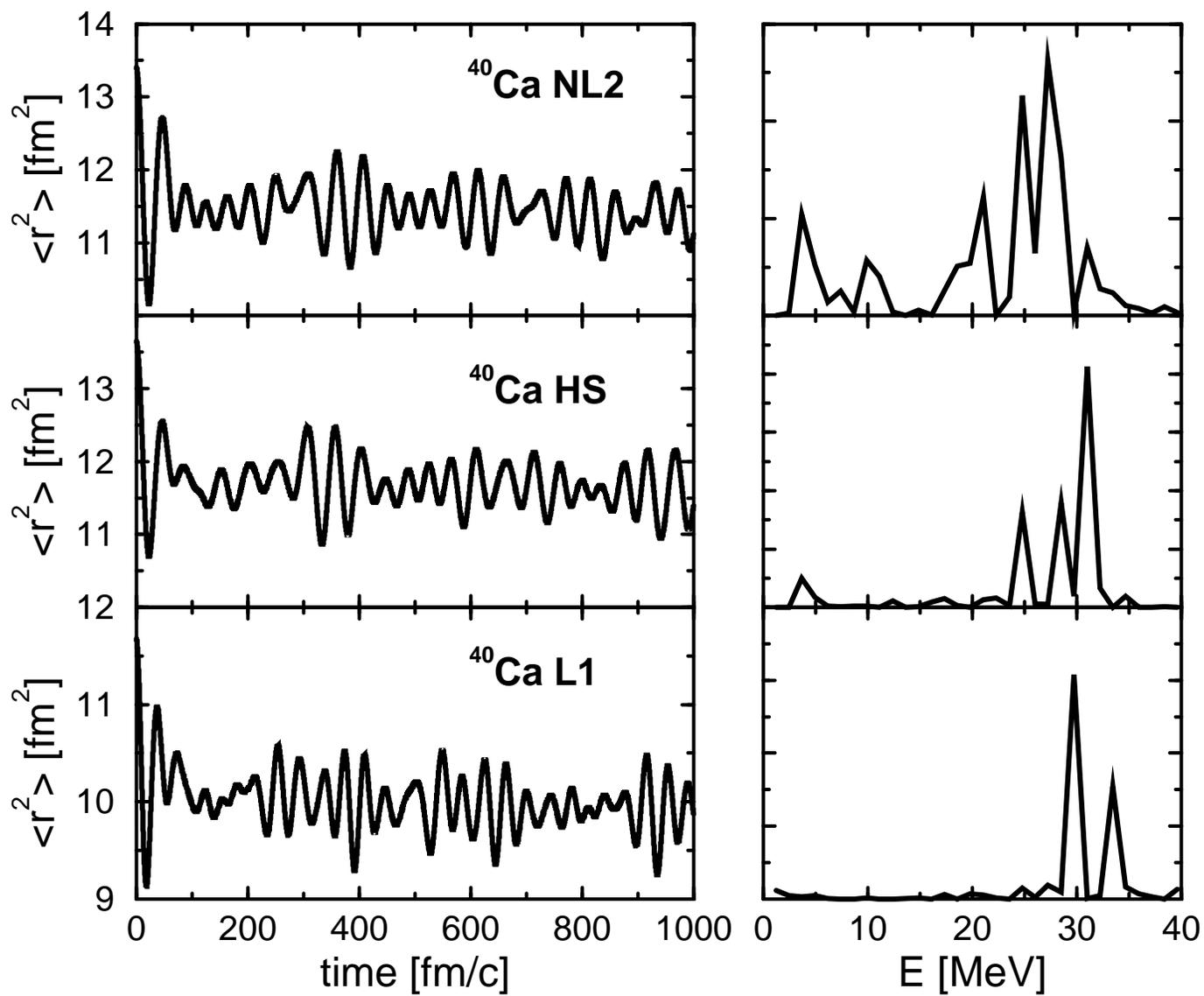
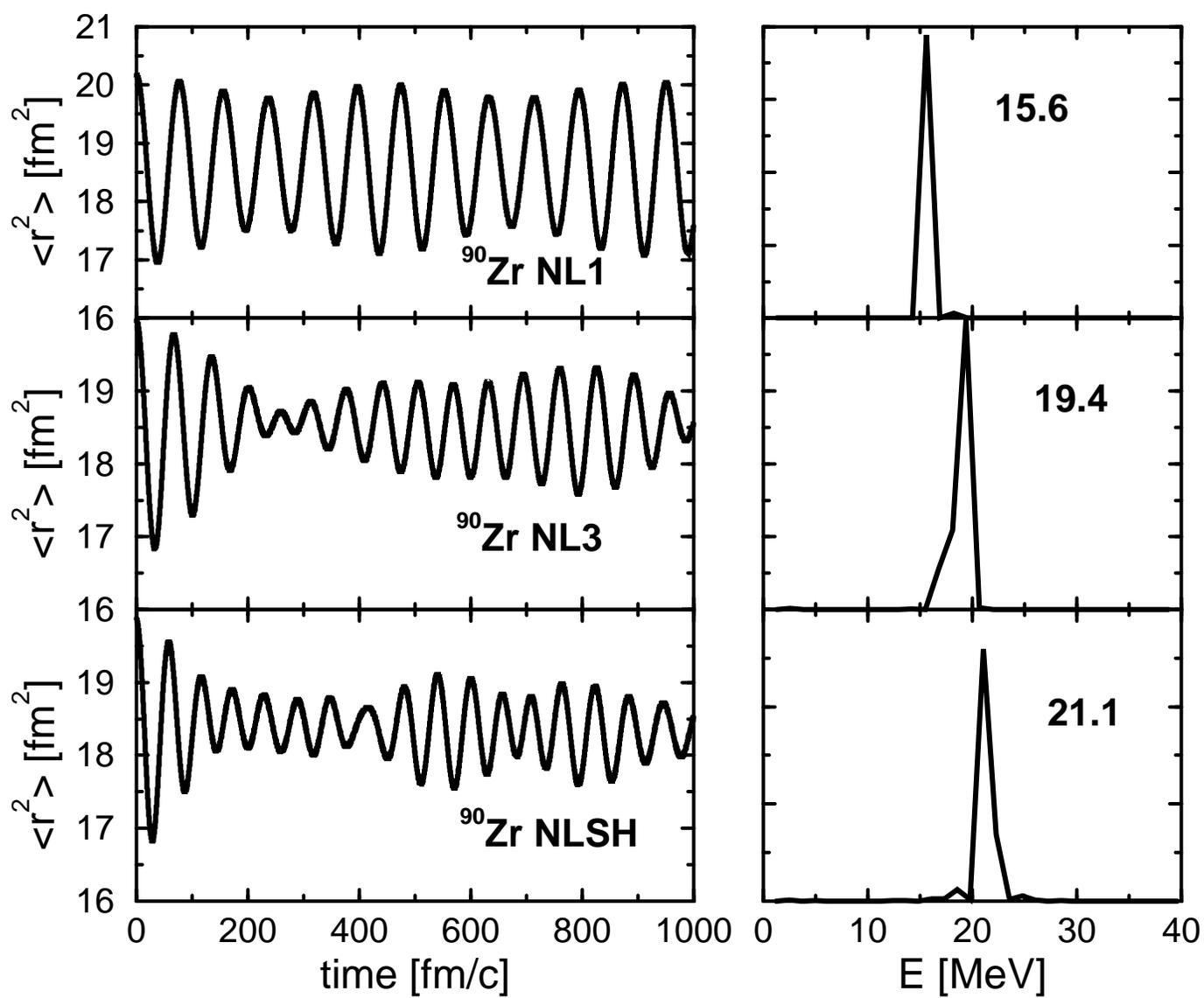
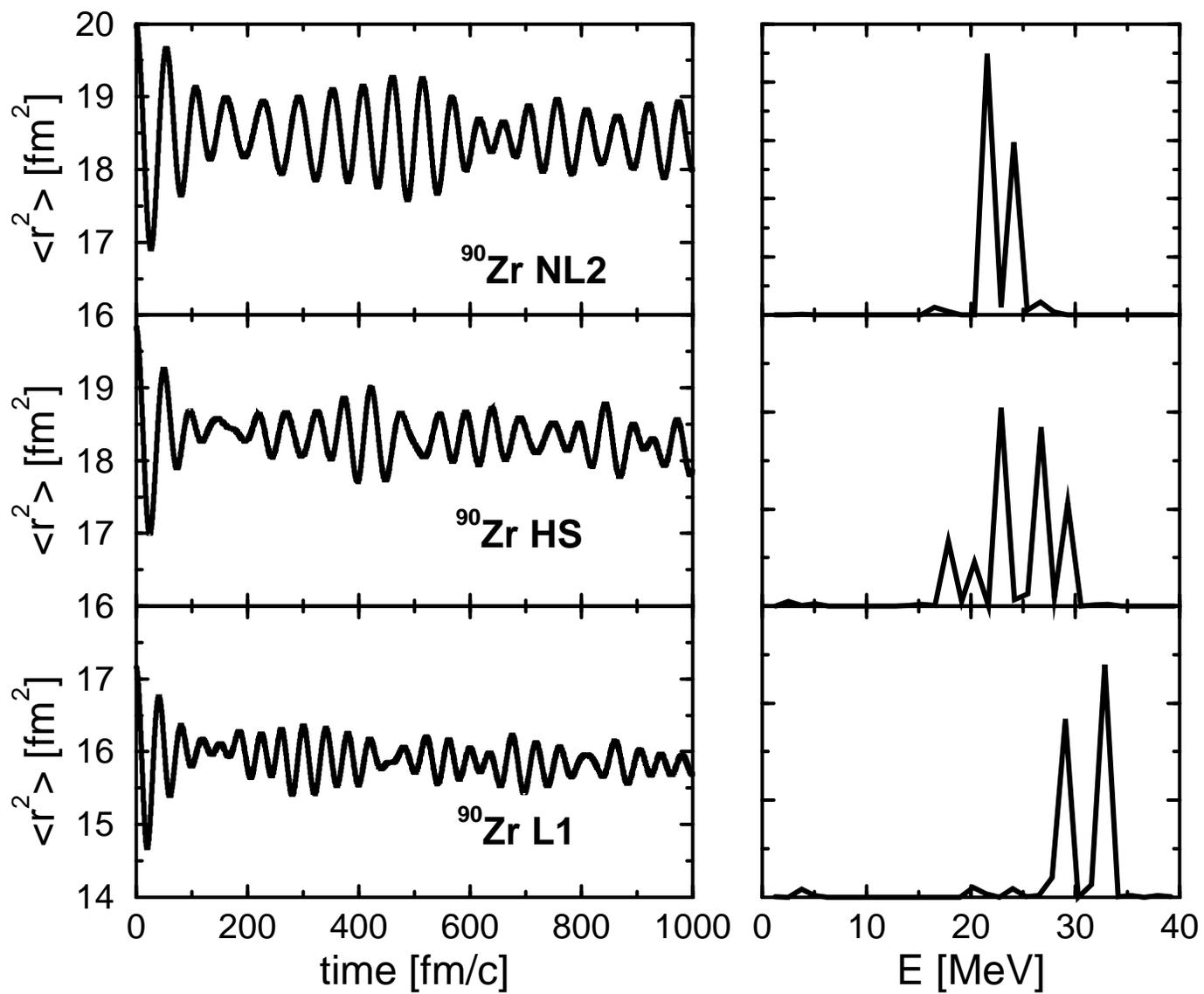
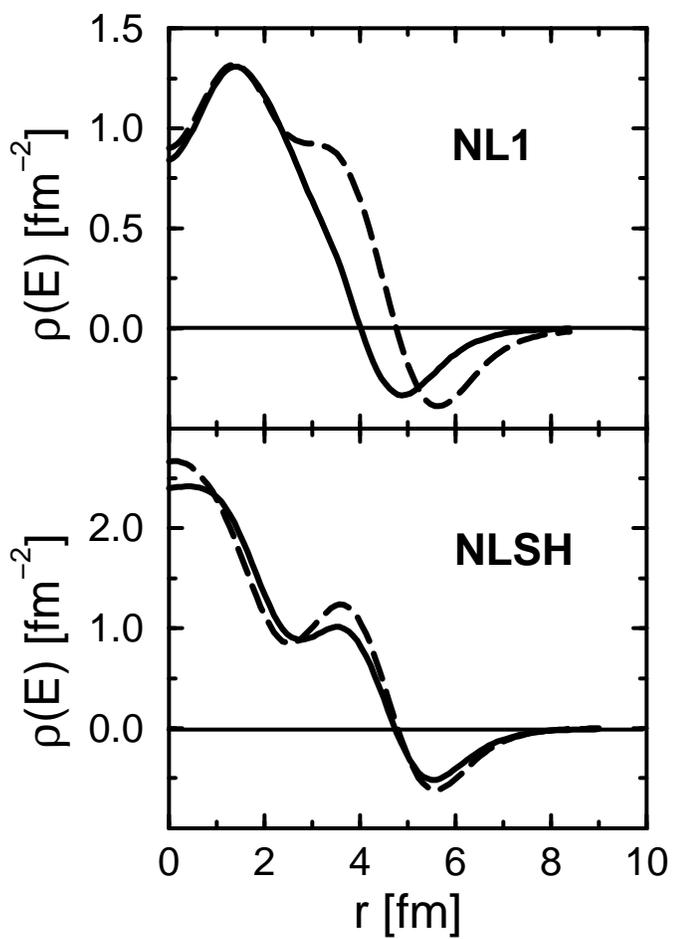
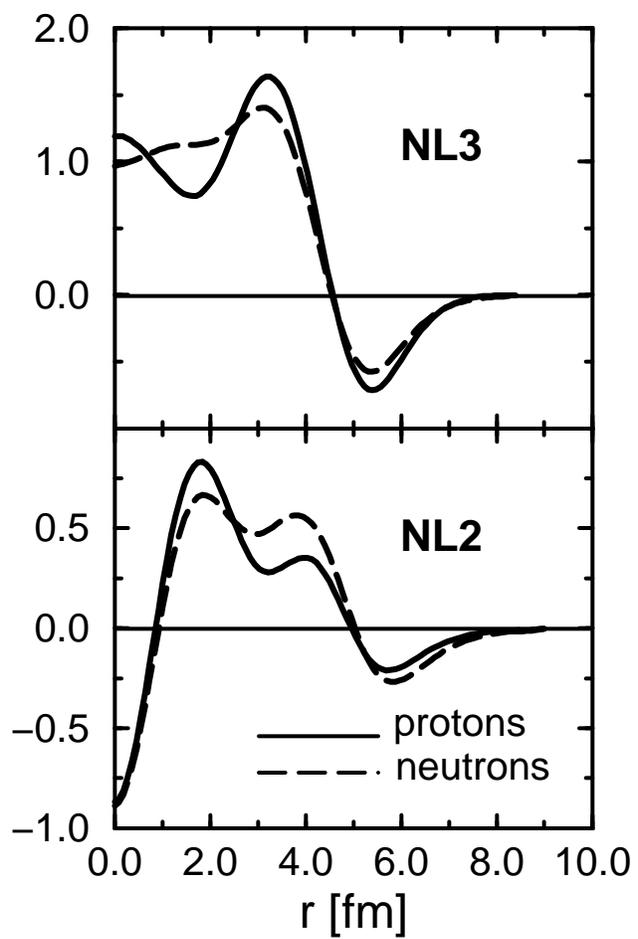
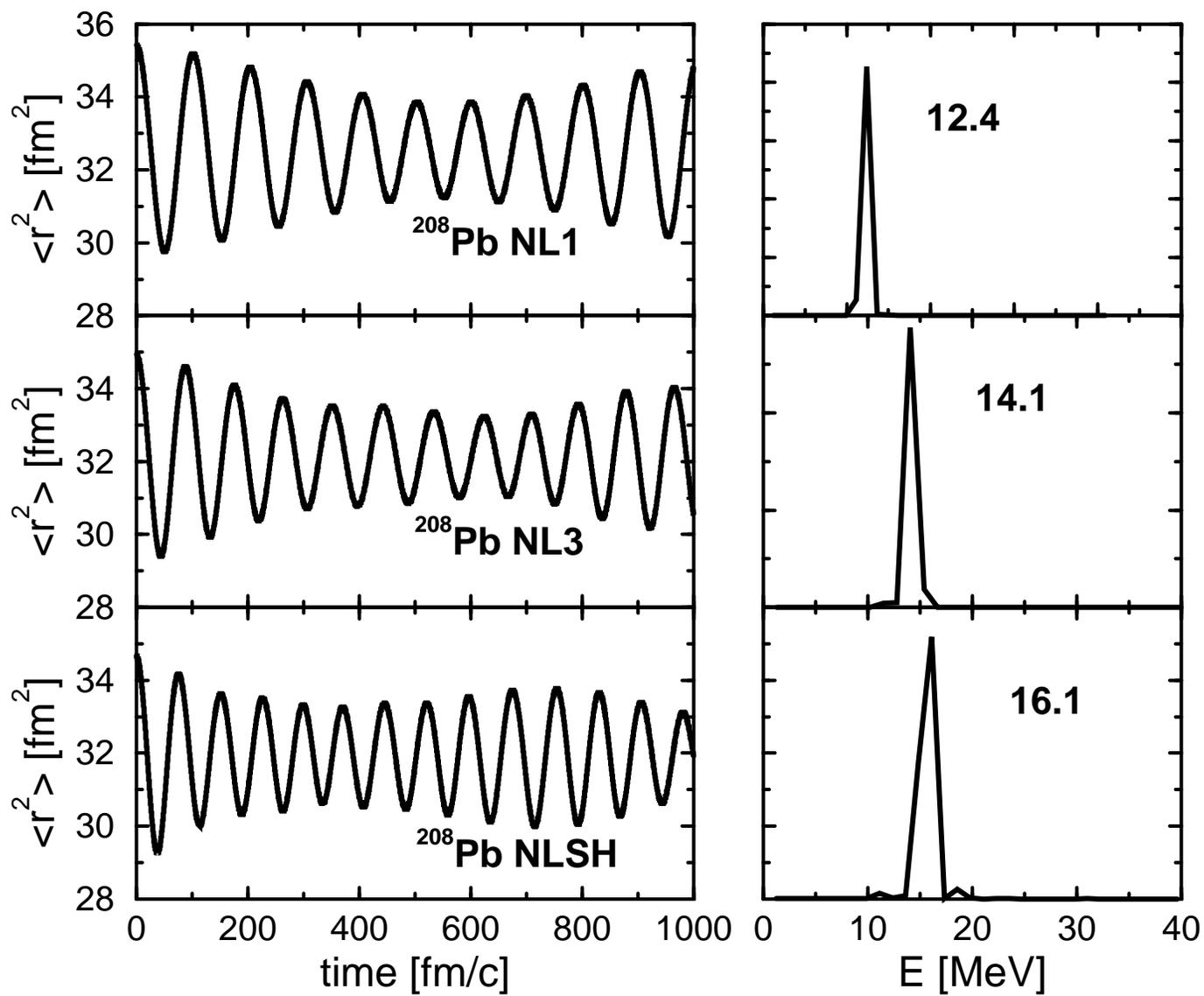
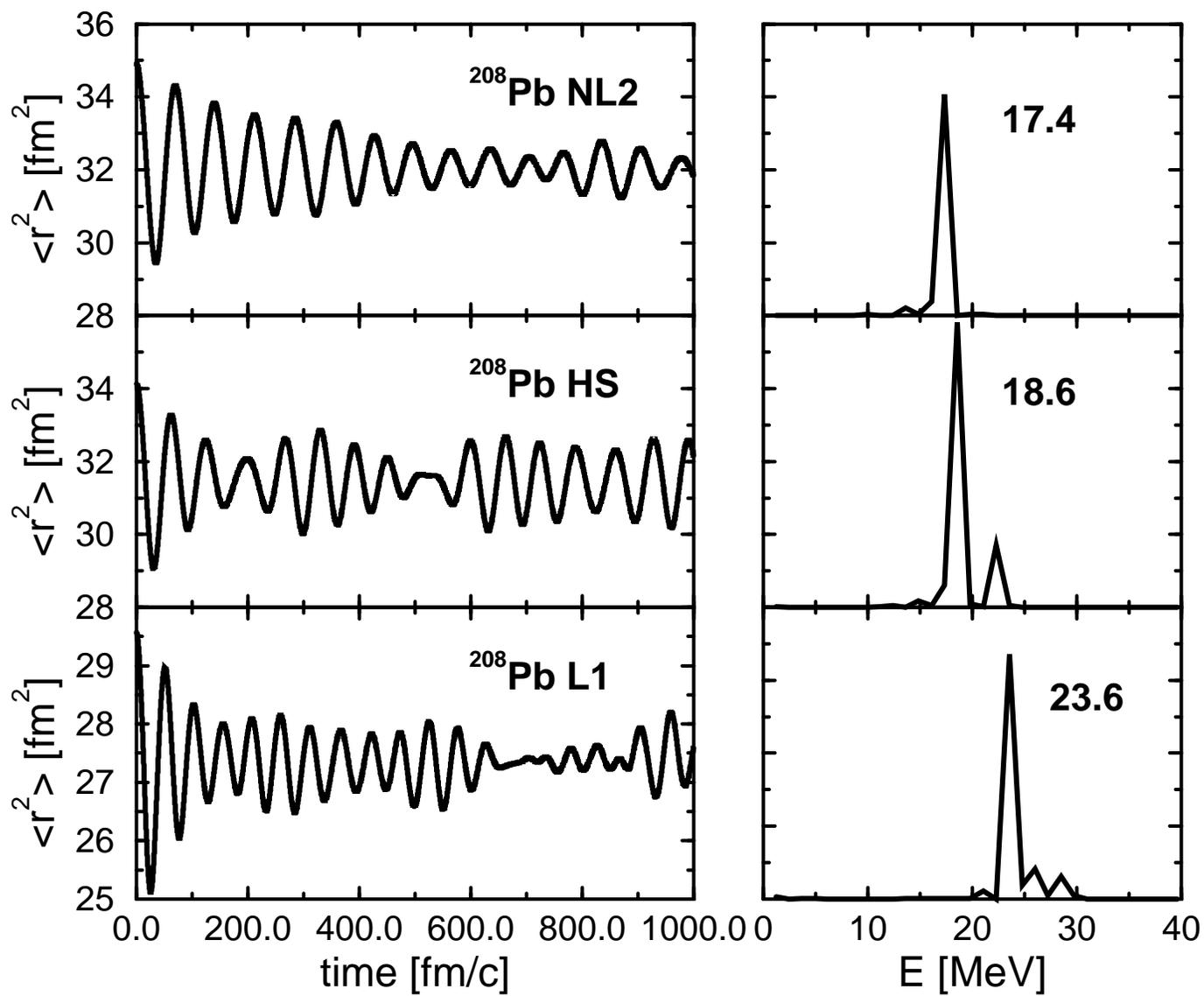
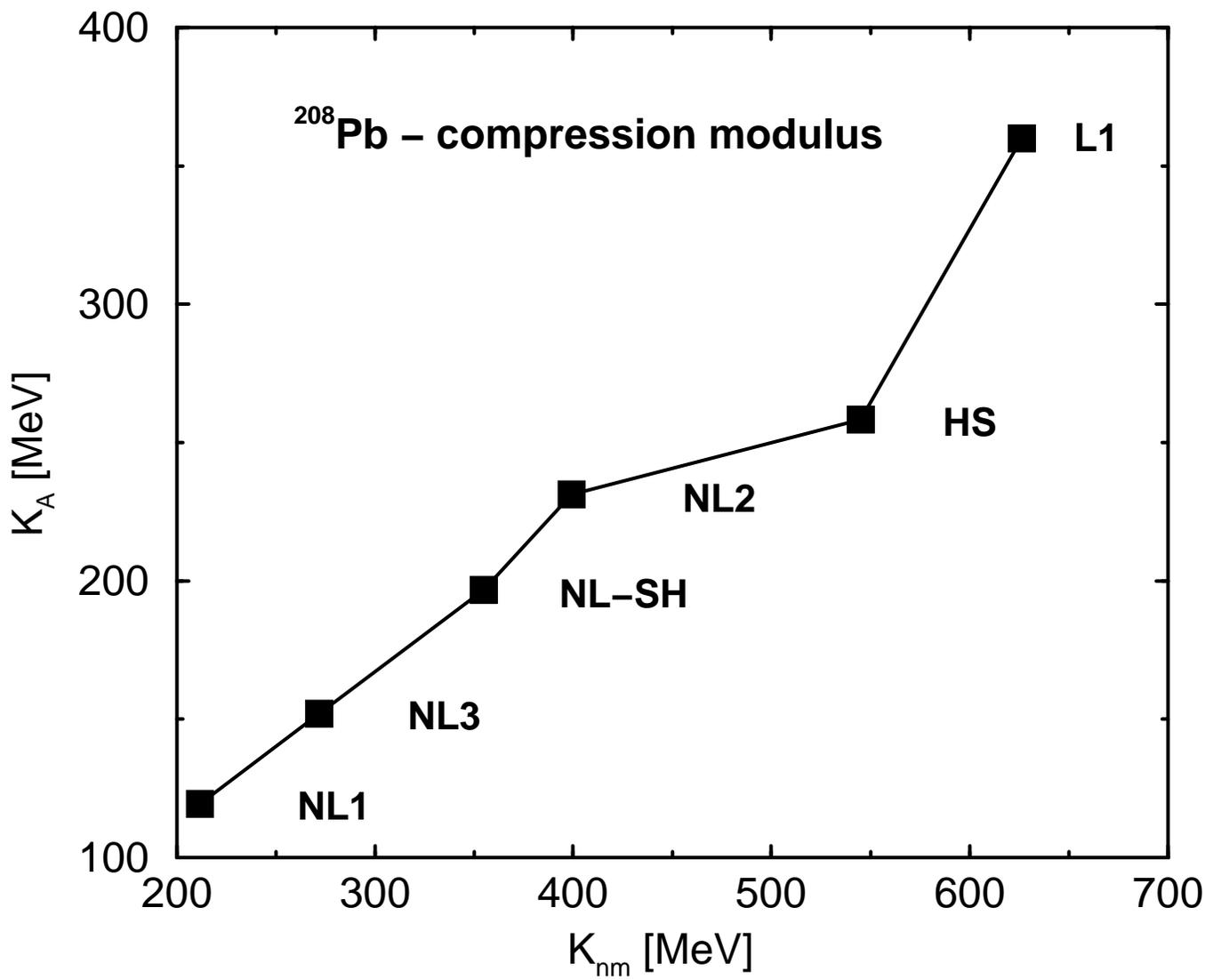
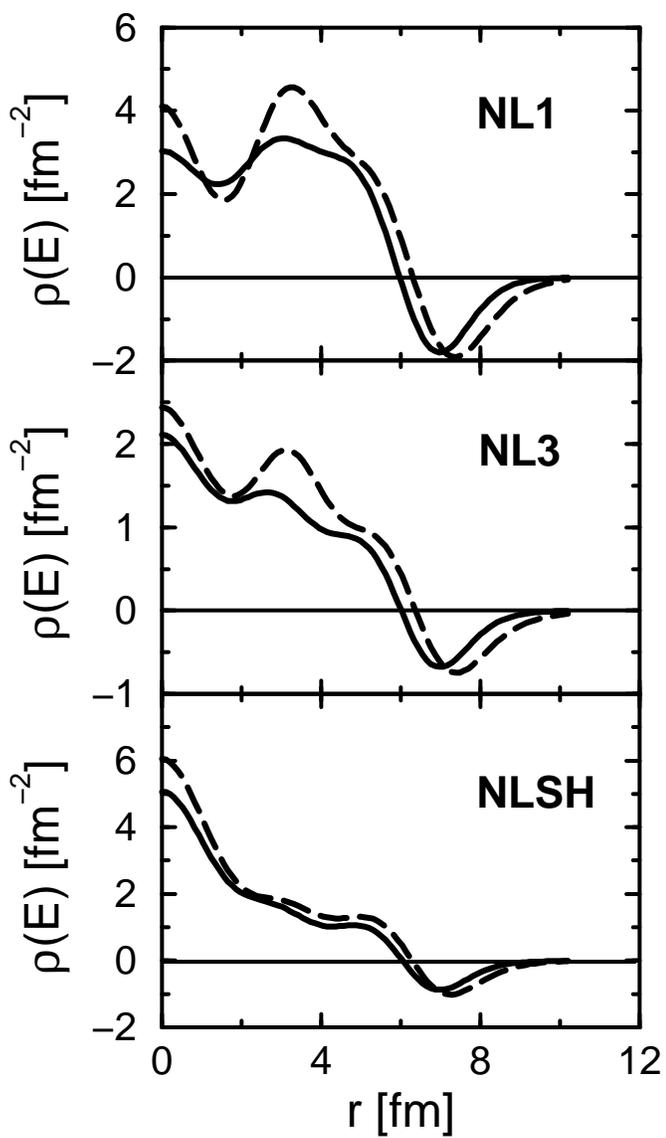
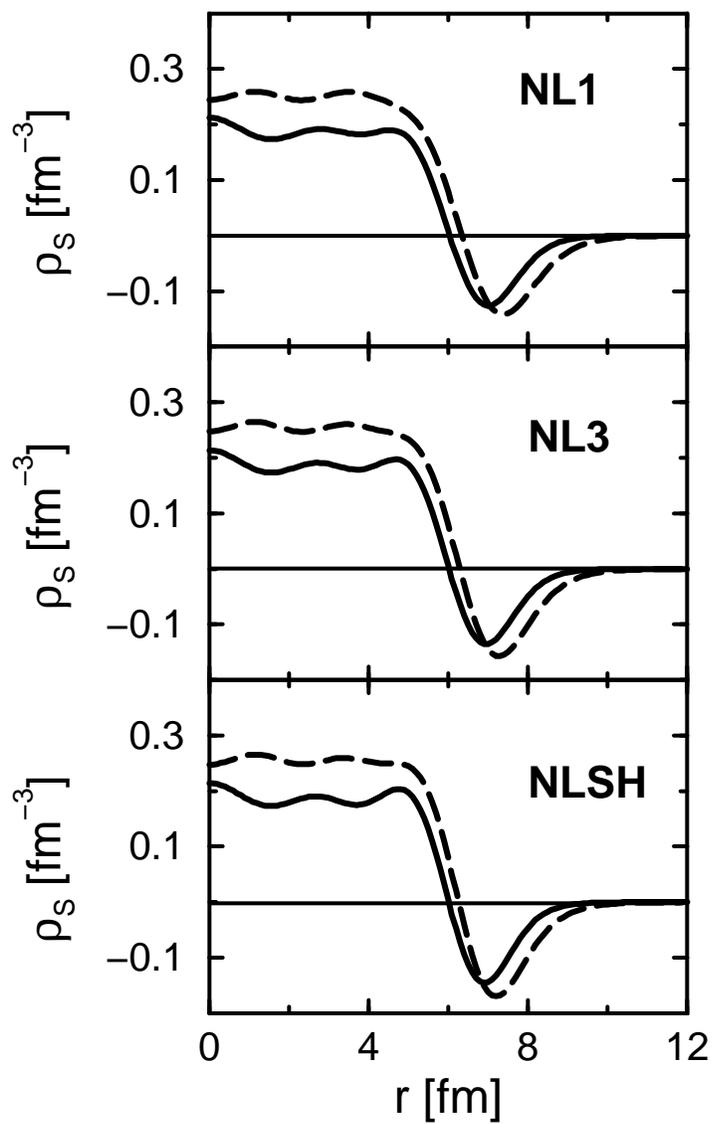
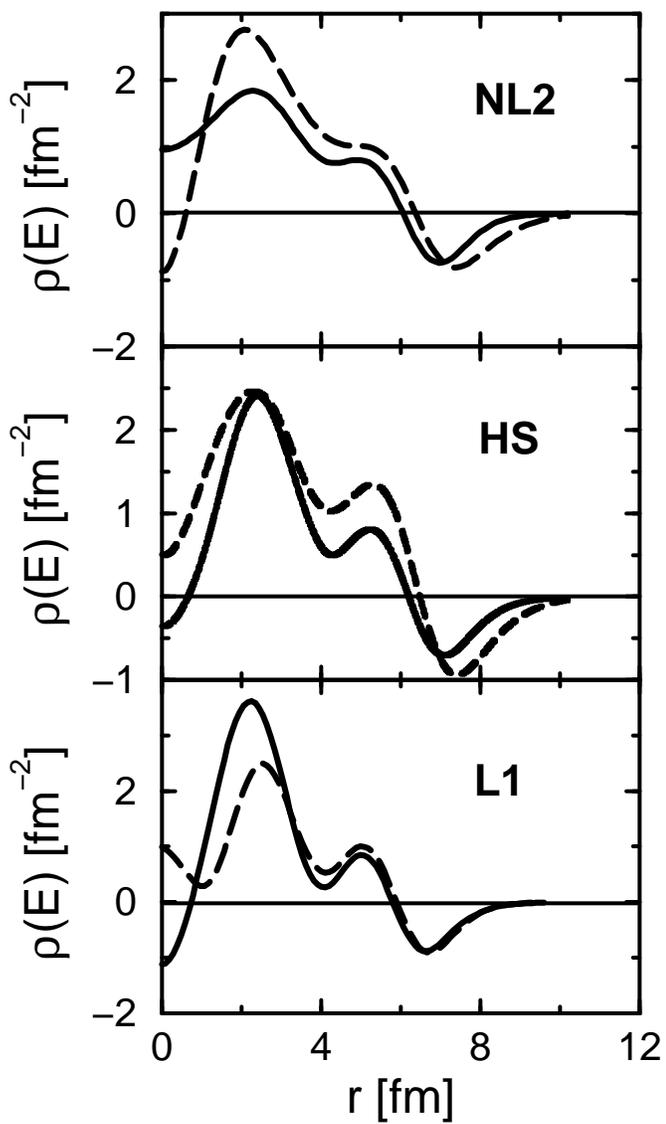
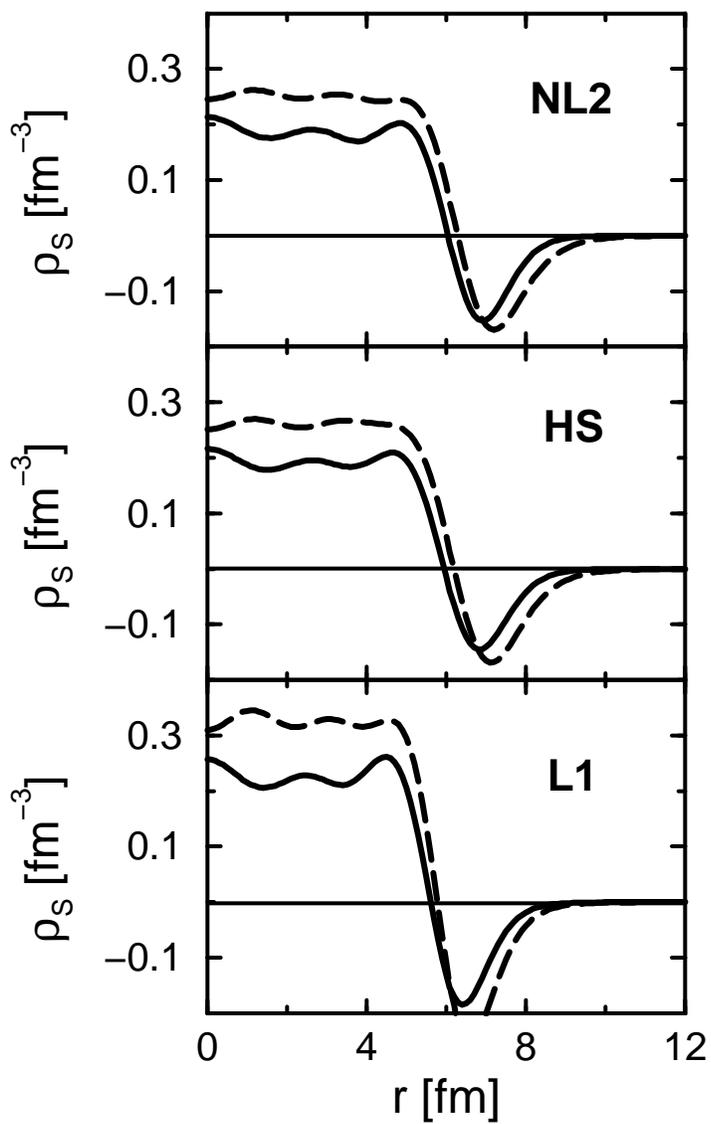
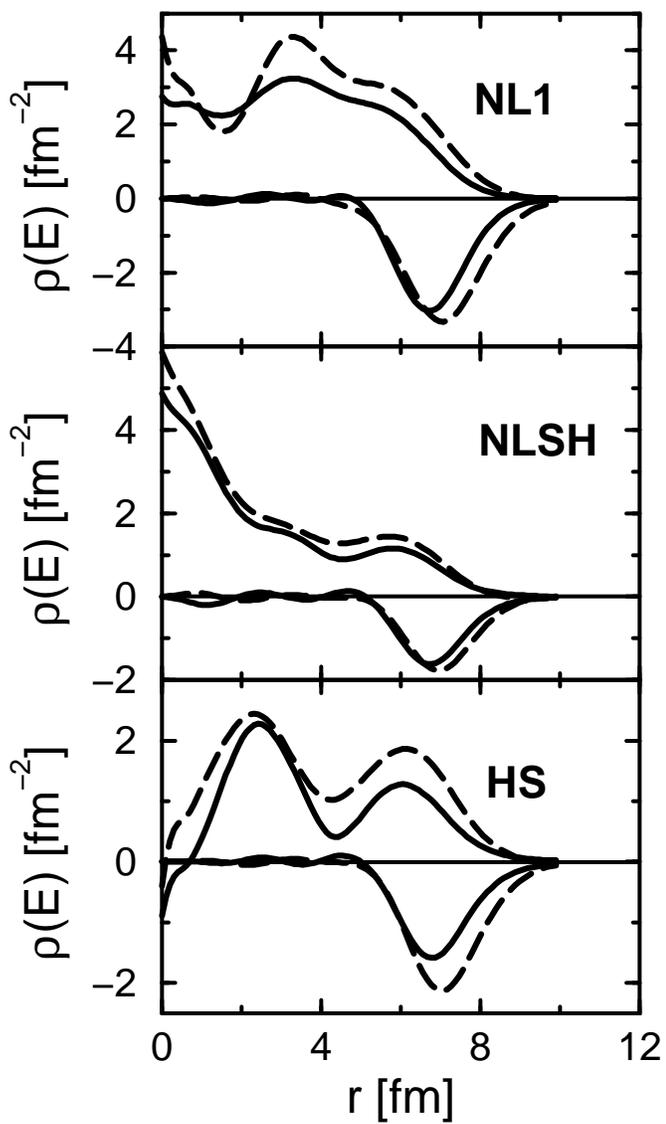
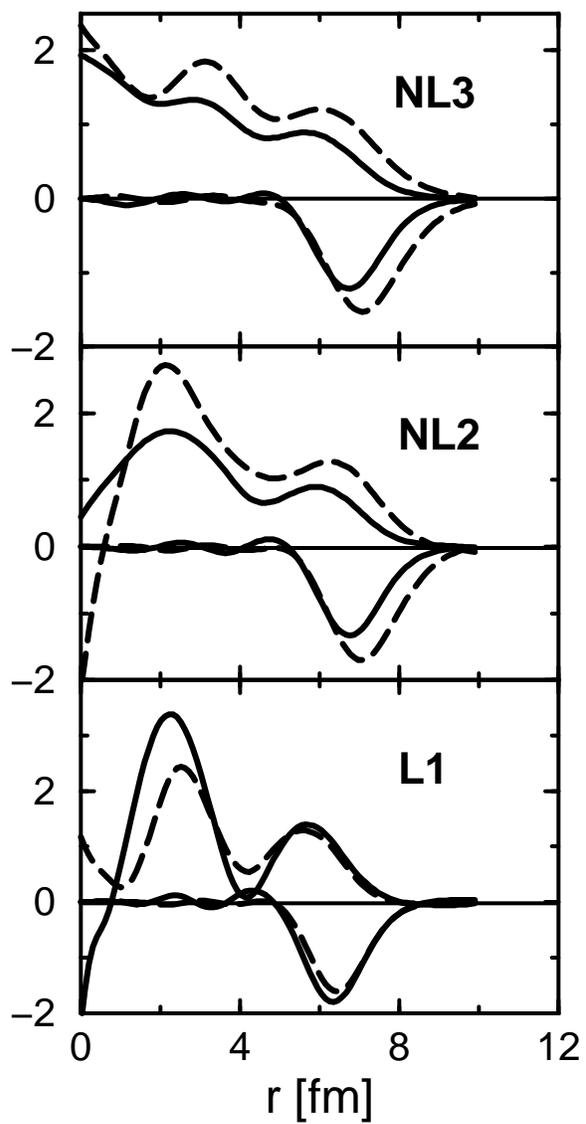
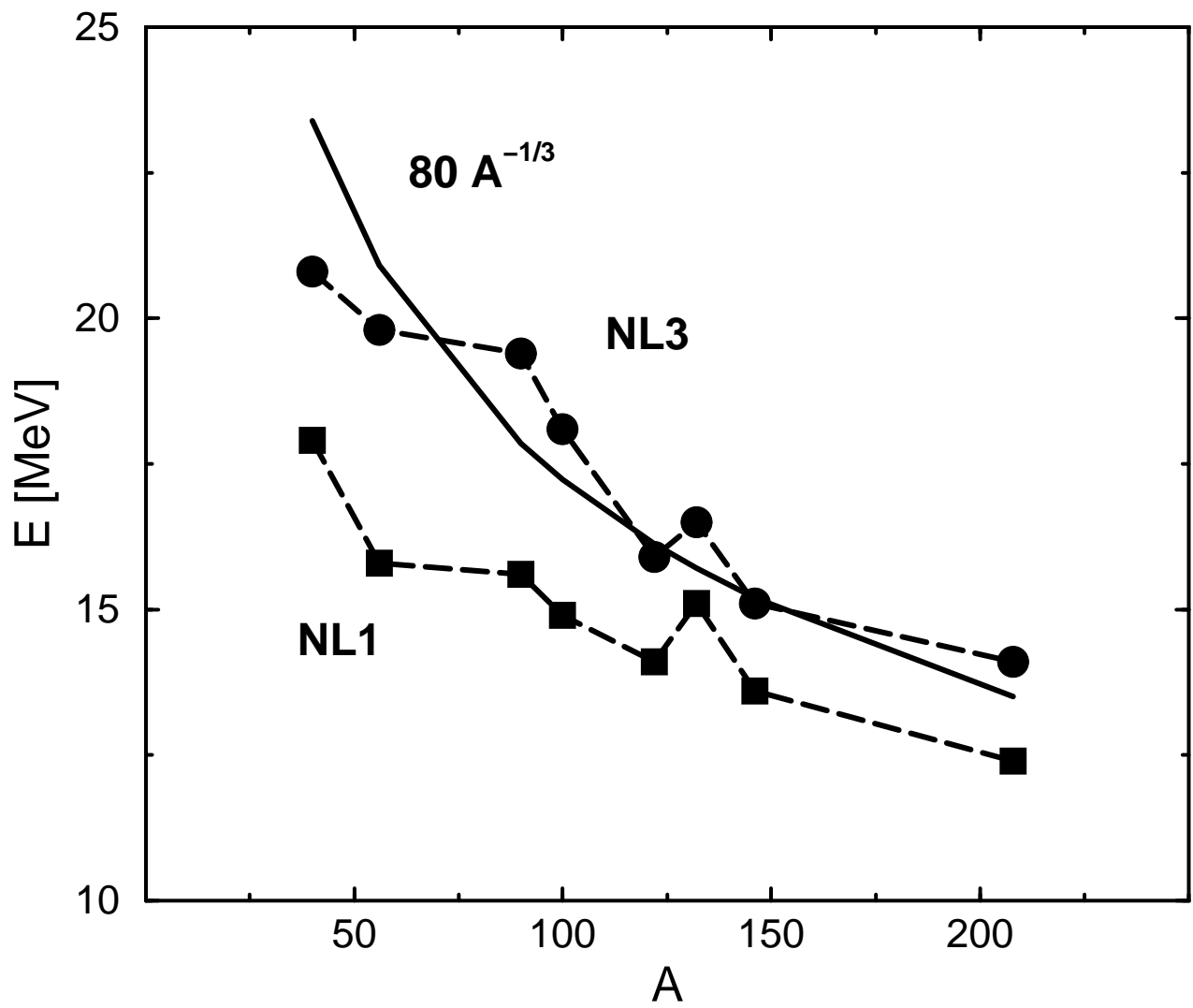
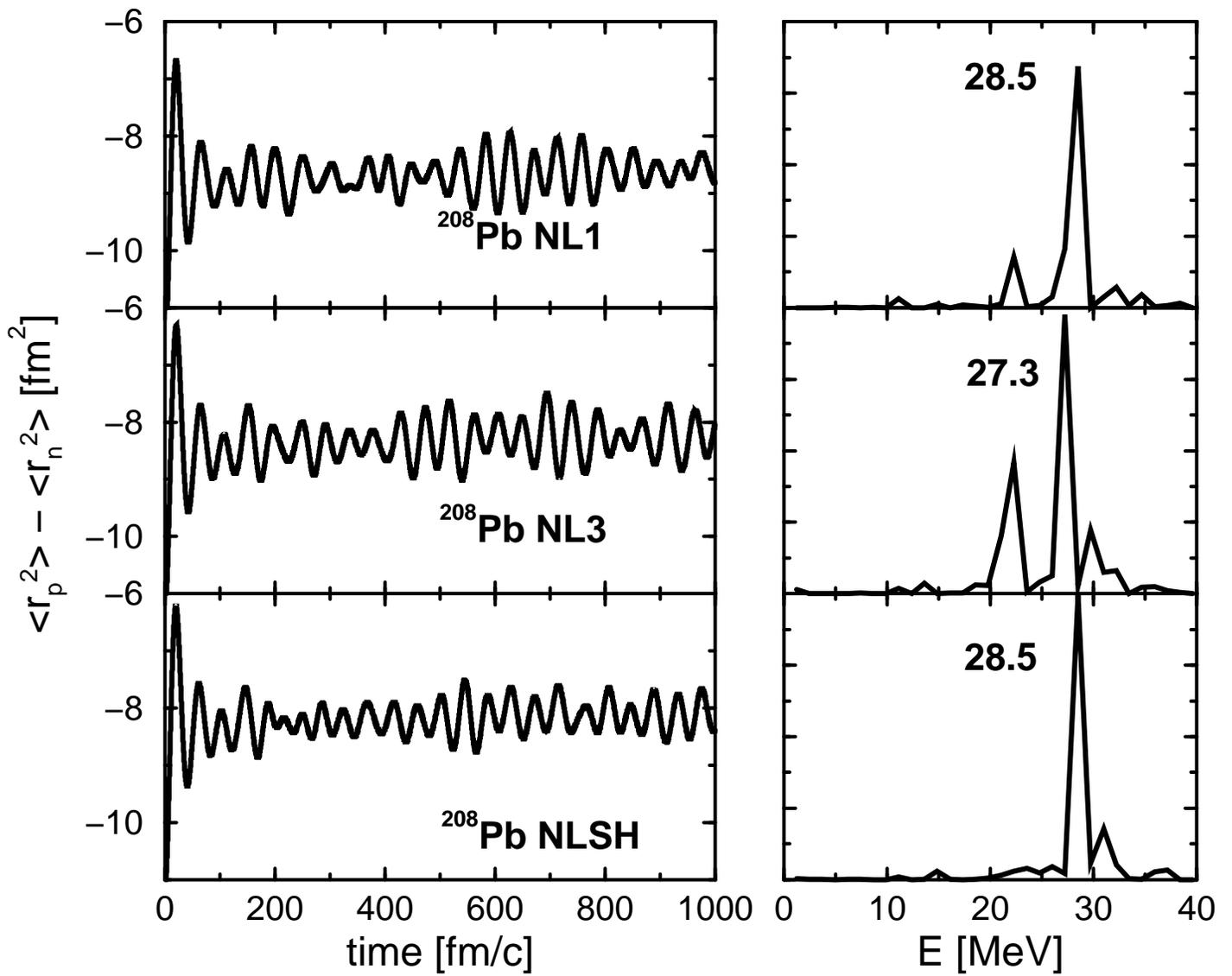
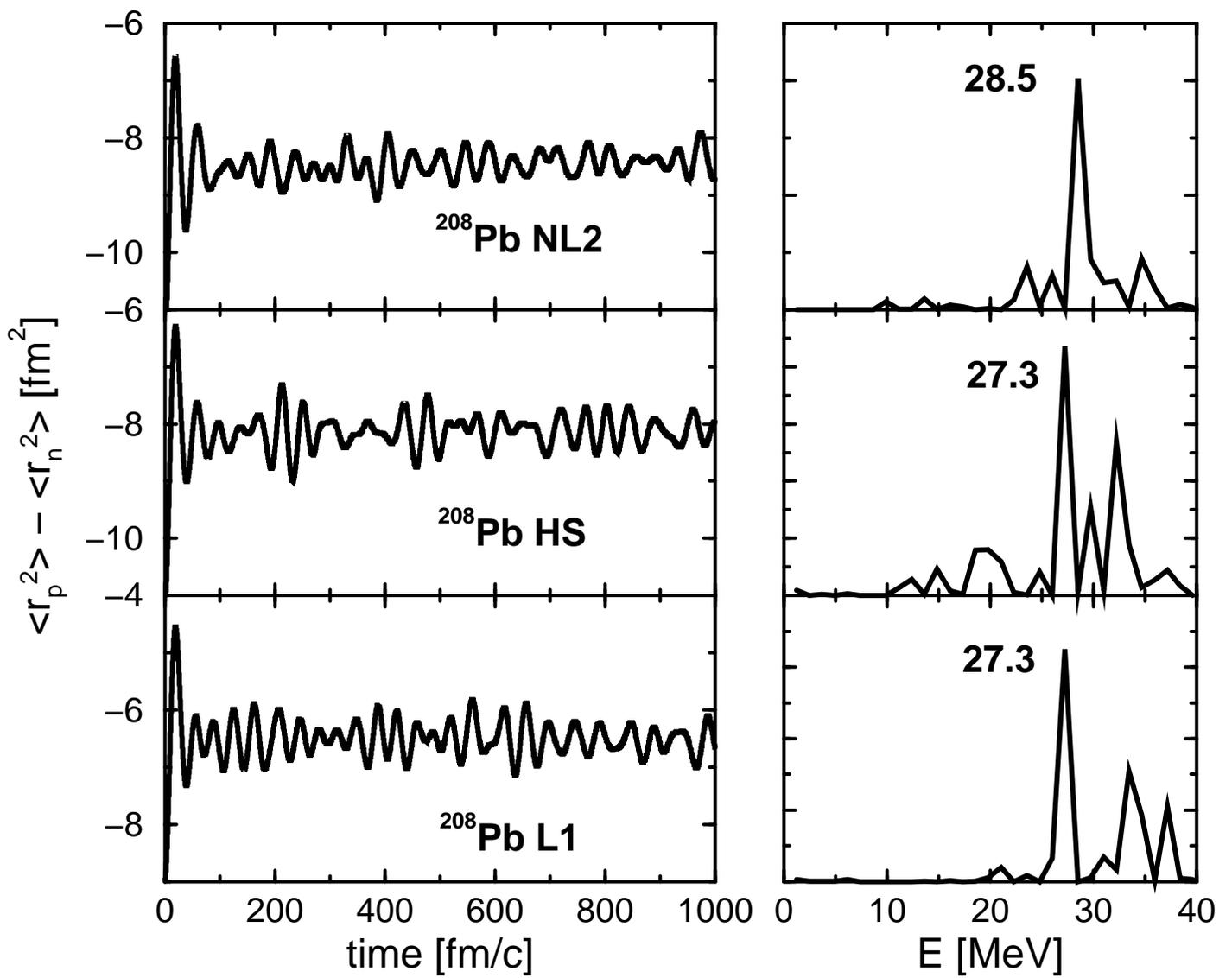
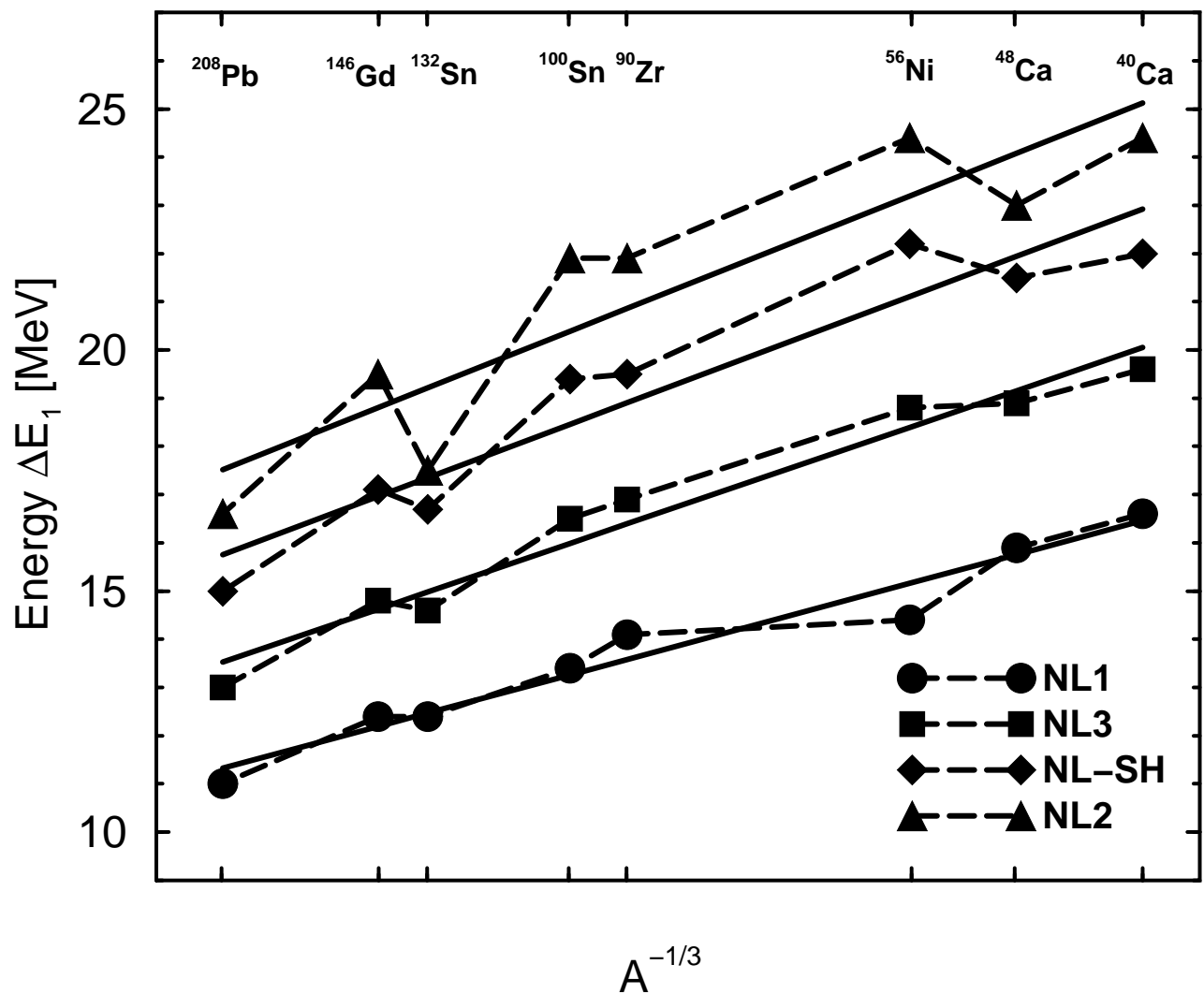
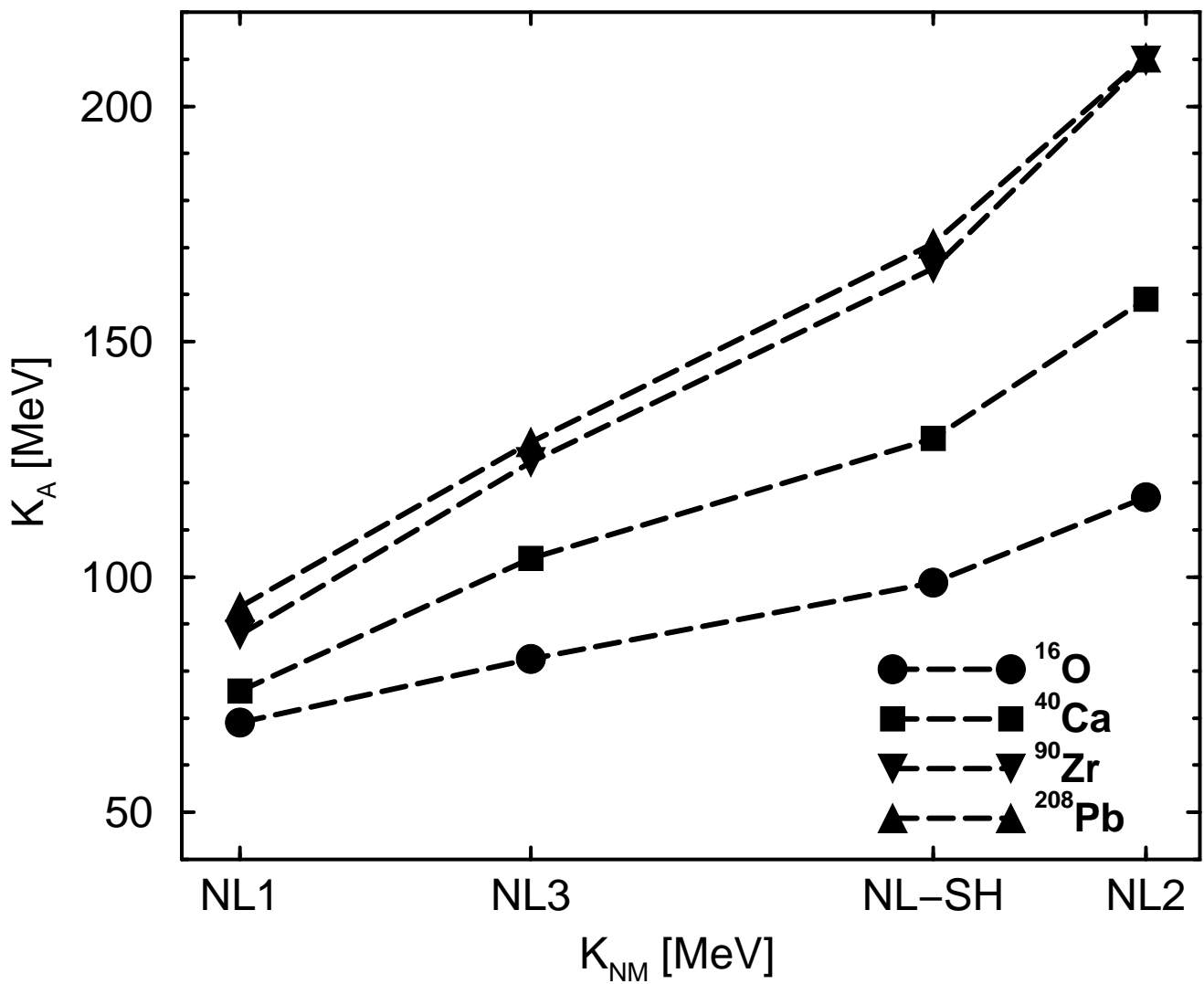
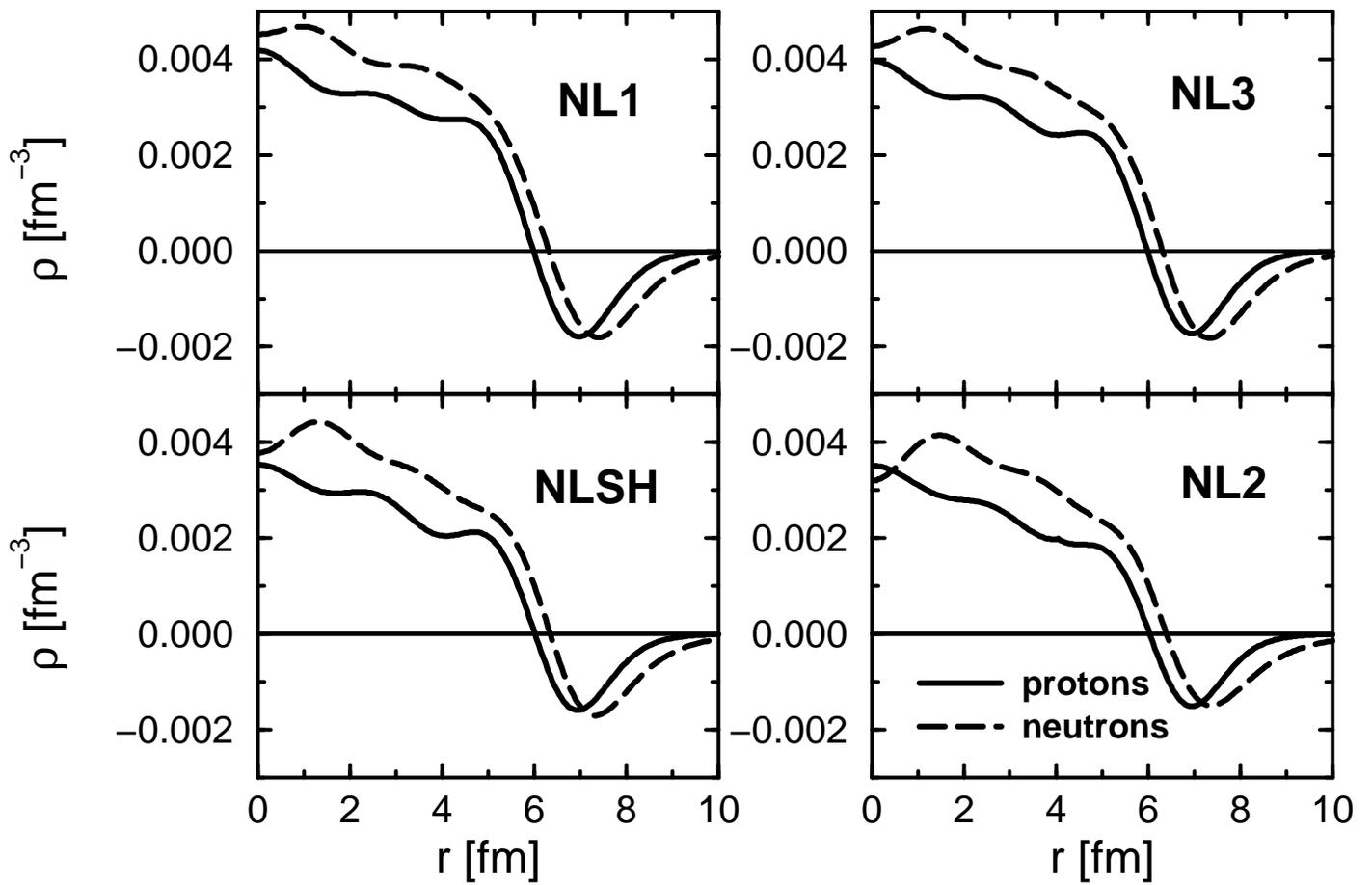
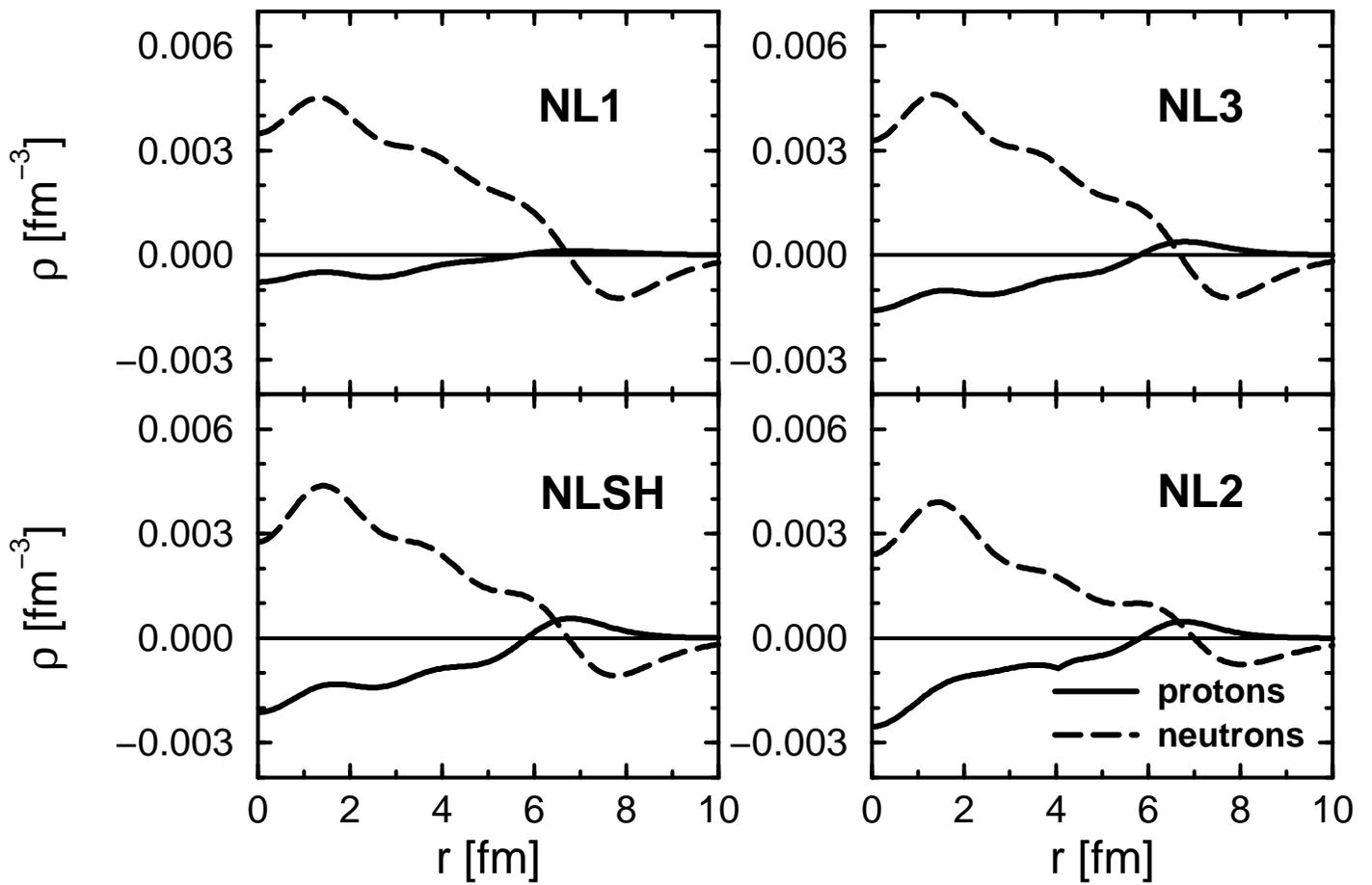
\end{document}